# Numerical Solution of Advection-Diffusion Equation Using Preconditionar as Incomplete LU Decomposition and the BiCGSTAB Aceleration Method

Dibakar Datta, Jacobo Carrasco Heres

Erasmus MSc in Computational Mechanics

Ecole Centrale de Nantes, FRANCE

Present Address: dibakar_datta@brown.edu  or  dibdatlab@gmail.com

**Abstract:** In the present study, an advection-diffusion problem has been considered for the numerical solution. The continuum equation is discretized using both upwind and centered scheme. The linear system is solved using the ILU preconditioned BiCGSTAB method. Both Dirichlet and Neumann boundary condition has been considered. The obtained results have been compared for different cases.

# 1. Introduction:

Most of the real life problems can be mathematically modeled with the Partial Differential Equation (PDE). A well-defined problem can be posed with the proper boundary condition based on the real life phenomenon. However, apart from few, most of the PDE has no close form solution. Hence the discretization of the continuum equation and numerical solution of the equation is very important. However, stability converges etc are the key and important issues as far as numerical solution is considered. Depending on the scheme, the solution may have oscillation or spike. Hence it will lead to the erroneous conclusions as compared to the real physics. Also it is required to optimize the CPU time. Large number of mesh size will give better result. But CPU time will be very high. The dimension of a real life problem is itself very high. Hence it is necessary to have stable and convergence solution with the optimized CPU time consumption. With this point of view, in the present problem, the ILU preconditioned system is solved using the BiCGSTAB method.

The report is organized in the following ways. Section 2 details the problem. Section 3 deals with the solution manufacturing. In section 4, discretization of the continuum equation and linearization of the system has been described in details. Section 5 deals with the details of different preconditioned and linear solver. The ILU and BiCGSTAB which are considered for the present problem has been detailed. Section 6 includes the Matlab codes. In Section 7, the obtained results have been analyzed in details. Finally the report ends with conclusive discussion in section 8.

# 2. Problem Statement:

To solve the advection-diffusion equation:

$$\varepsilon\left(\frac{\partial^2 u}{\partial x^2}+\frac{\partial^2 u}{\partial y^2}\right)+a(x,y)\frac{\partial u}{\partial x}+b(x,y)\frac{\partial u}{\partial y}=f(x,y) \quad \ldots (1)$$

$\varepsilon > 0$ ; $x \in [0, X]$ and $y \in [0, Y]$  $a(x,y) = 1 + x^2$ ; $b(x,y) = X\bar{e}^y$

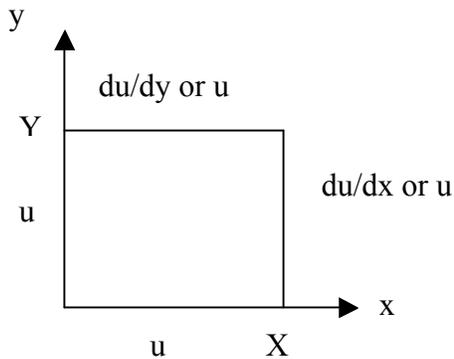

Always Dirichlet on x=0, y=0;

Either Neumann at y=Y and x=X

Or Dirichlet on y=Y and x=X

**Question:** Compute f(x,y) such that equation (1) is satisfied. Given that $u(x,y) = e^{-x/X}\left(1 - e^{-y/Y}\right)y$

*Question:*

Solve the equation after differencing the equation

- Central Differencing

- Upwind Differencing

Solve the linear system by the preconditioned (Incomplete LU Decomposition) BiCGSTAB method.

- Plot the residual reduction (L$_2$ norm and the maximum norm).
- Plot the error map
- Estimate the truncation error.
- Plot the smoothing factor of the iteration.

# 3. Solution Manufacturing:

Given $u(x,y) = e^{-x/X}\left(1-e^{-y/Y}\right)y$

Hence:

$$\frac{\partial u}{\partial x} = \frac{-1}{X}e^{-x/X}\left(1-e^{-y/Y}\right)y \quad ; \quad \frac{\partial^2 u}{\partial x^2} = \frac{1}{X^2}e^{-x/X}\left(1-e^{-y/Y}\right)y$$

$$\frac{\partial u}{\partial y} = e^{-x/X}\left(1-e^{-y/Y}\right) + \frac{y}{Y}e^{-\left(\frac{x}{X}+\frac{y}{Y}\right)} \quad ; \quad \frac{\partial^2 u}{\partial y^2} = \left(\frac{2}{Y}-\frac{y}{Y^2}\right)e^{-\left(\frac{x}{X}+\frac{y}{Y}\right)}$$

Substituting in the equation (1), we get:

a(x,y) = 1 + x2

a(x,y) ∂u∂x = (1 + x2 ) ( - 1X e-xX ( 1 - e-yY )y)

b(x,y) ∂u∂y = X e-y (e-xX ( 1- e-yY ) + yY e-xX e-yY ) = X e-y (e-xX- e-xX+yY + yY e-xX+yY )

a(x,y) ∂u∂x+ (x,y) ∂u∂y = (1 + x2 ) ( - 1X e-xX ( 1 - e-yY )y) + X e-y (e-xX- e-xX+yY + yY e-xX+yY ) = - y ( 1+ x2)X e-xX + y ( 1+ x2)X e-xX+yY + X e-y e-xX - X e-y e-xX+yY + +XYye-ye-xX+yY = e-xX (X e-y - y ( 1+ x2)X) + e-xX+yY(y ( 1+ x2)X - Xe-y + XYy e-y )

f(x,y) = ε ( yX2 e-xX + e-xX+yY( 2Y- yY2- yX2)) + + e-xX (Xe-y- y ( 1+ x2)X ) + e-xX+yY (y ( 1+ x2)X - Xe-y +XYy e-y)

# 4. Discretization of the Continuum equation

For the present problem, two types of discretization scheme have been compared for the comparison of the solution- centered and hybrid.

∂2u∂x2 ]$_{i,j}$ = Ui+1,j - 2 Ui,j + Ui-1,j∆x2+ O∆x2

∂2u∂y2 ]$_{i,j}$ = Ui,j+1 - 2 Ui,j + Ui,j-1∆y2+ O∆y2

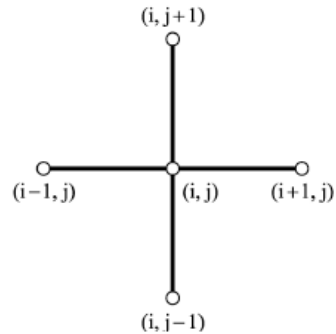

**Hybrid Scheme:**

∂u∂x ]$_{i,j}$ = Ui,j−Ui−1,j ∆x + O∆x

∂u∂x ]$_{i,j}$ = Ui,j−Ui,j−1 ∆y + O∆y

ε [ Ui+1,j − 2 Ui,j + Ui−1,j∆x2 + Ui,j+1 − 2 Ui,j + Ui,j−1∆y2 ] + aij Ui,j−Ui−1,j ∆x + bij Ui,j−Ui,j−1 ∆y= fij

or

ε [ (Ui+1,j − 2 Ui,j + Ui−1,j) ∆y2 + (Ui,j+1 − 2 Ui,j + Ui,j−1 ) ∆x2 ] +

$+ a_{ij}(U_{i,j} - U_{i-1,j})\Delta x \Delta y^2 + b_{ij}(U_{i,j} - U_{i,j-1})\Delta x^2 \Delta y = f_{ij}\Delta x^2 \Delta y^2$

or

$[\varepsilon \Delta x^2 - b_{ij}\Delta x^2 \Delta y] U_{i,j-1} + [\varepsilon \Delta y^2 - a_{ij}\Delta x \Delta y^2] U_{i-1,j} + [-2\varepsilon \Delta y^2 - 2\varepsilon \Delta x^2 + a_{ij}\Delta x \Delta y^2 + b_{ij}\Delta x^2 \Delta y] U_{i,j} + [\varepsilon \Delta y^2] U_{i+1,j} + [[\varepsilon \Delta x^2] U_{i,j+1} = f_{ij}\Delta x^2 \Delta y^2$

$a_{ij} = 1 + x_i^2$

$b_{ij} = X\, e^{-y_j}$

**Centered Scheme:**

$\partial u / \partial x \,]_{i,j} = \dfrac{U_{i+1,j} - U_{i-1,j}}{2\Delta x} + O\Delta x$

$\partial u / \partial x \,]_{i,j} = \dfrac{U_{i,j+1} - U_{i,j-1}}{2\Delta y} + O\Delta y$

$\varepsilon\left[\dfrac{U_{i+1,j} - 2U_{i,j} + U_{i-1,j}}{\Delta x^2} + \dfrac{U_{i,j+1} - 2U_{i,j} + U_{i,j-1}}{\Delta y^2}\right] + a_{ij}\dfrac{U_{i+1,j} - U_{i-1,j}}{2\Delta x} + b_{ij}\dfrac{U_{i,j+1} - U_{i,j-1}}{2\Delta y} = f_{ij}$

or

$\varepsilon[(U_{i+1,j} - 2U_{i,j} + U_{i-1,j})\Delta y^2 + (U_{i,j+1} - 2U_{i,j} + U_{i,j-1})\Delta x^2] +$

$+ 0.50\, a_{ij}(U_{i+1,j} - U_{i-1,j})\Delta x \Delta y^2 + 0.50\, b_{ij}(U_{i,j+1} - U_{i,j-1})\Delta x^2 \Delta y = f_{ij}\Delta x^2 \Delta y^2$

or

$[\varepsilon \Delta x^2 - 0.50\, b_{ij}\Delta x^2 \Delta y] U_{i,j-1} + [\varepsilon \Delta y^2 - 0.50\, a_{ij}\Delta x \Delta y^2] U_{i-1,j} + [-2\varepsilon \Delta y^2 - 2\varepsilon \Delta x^2] U_{i,j} + [\varepsilon \Delta y^2 + 0.50\, a_{ij}\Delta x \Delta y^2] U_{i+1,j} + [[\varepsilon \Delta x^2 + 0.50\, b_{ij}\Delta x^2 \Delta y] U_{i,j+1} = f_{ij}\Delta x^2 \Delta y^2$

# 5. Overview of Preconditioning Krylov Subspace methods for solving the linear system:

## 5.1: Preconditioning:

**Table 5.1:** Different preconditioning algorithm:

| Preconditioner Algorithm | PCType | Matrix types* | External Package | Parallel | Complex |
|---|---|---|---|---|---|
| Jacobi | PCJACOBI | aij,baij,sbaij,dense | | X | X |
| Point Block Jacobi | PCPBJACOBI | baij,bs=2,3,4,5 | | X | X |
| SOR | PCSOR | seqdense,seqaij,seqsbaij,mpiaij | | | X |
| Point Block SOR | | seqbaij,bs=2,3,4,5 | | | X |
| Block Jacobi | PCBJACOBI | aij,baij,sbaij | | X | X |
| Additive Schwarz | PCASM | aij,baij,sbaij | | X | X |
| ILU(k) | PCILU/PCICC | seqaij,seqbaij | | | X |
| ICC(k) | | seqaij,seqbaij | | | X |
| ILU dt | | seqaij | Sparsekit | | |
| ILU(0)/ICC(0) | | aij | BlockSolve95 | X | |
| ILU(k) | PCHYPRE | aij | Euclid/HyPre | X | |
| ILU dt | | aij | Euclid/HyPre | X | |
| Matrix-free | PCSHELL | | | X | X |
| Multigrid/infrastructure | PCMG | | | X | X |
| Multigrid/geometric structured grid | DMMG | | | X | X |
| Multigrid algebraic | PCHYPRE | aij | BoomerAMG/HyPre | X | |
| | PCML | aij | ML/Trilinos | X | |
| | PC | baij | Prometheus | X | |
| Approximate inverses | PCHYPRE | aij | Parasails/HyPre | X | |
| | PCSPAI | aij | SPAI | X | |
| Balancing Neumann-Neumann | PCNN | is | | X | X |
| Direct solver | | | | | |
| LU | PCLU | seqaij,seqbaij | | | X |
| | | seqaij | MATLAB | | X |
| | | aij | Spooles | X | X |
| | | aij | PastuiX | X | X |
| | | aij | SuperLU, Sequential/Parallel | X | X |
| | | aij | MUMPS | X | X |
| | | seqaij | ESSL | | |
| | | seqaij | UMFPACK | | |
| | | dense | PLAPACK | X | X |
| Cholesky | PCCHOLESKY | seqaij,seqbaij | | | X |
| | | sbaij | Spooles | X | X |
| | | sbaij | PastuiX | X | X |
| | | sbaij | MUMPS | X | X |
| | | seqsbaij | DSCPACK | X | |
| | | dense | PLAPACK | X | X |
| | | matlab | MATLAB | | |
| | | aij | | X | |
| QR | | matlab | MATLAB | | |
| XXt and XYt | | aij | | X | |

* Matrix types

aij -	A matrix type to be used for sparse matrix

baij -	A matrix types to be used for block sparse matrix

sbaij -	A matrix type to be used for symmetric block sparse matrices

seqaij -	A matrix type to be used for sequential sparse matrices, based on compressed sparse row format.

mpiaij -    A matrix type to be used for parallel sparse matrices.
seqbaij -   A matrix type to be used for sequential block sparse matrices, based on block sparse compressed row format.
seqsbaij -  A matrix type to be used for sequential symmetric block sparse matrices, based on block compressed sparse row format.
dense -     A matrix type to be used for dense matrices.
seqdense -  A matrix type to be used for sequential dense matrices.
is -        A matrix type to be used for using the Neumann-Neumann type preconditioners.

## Incompleted LU

The Incomplete LU factorization technique with no fill-in, denoted by ILU (0), consists of taking the zero patterns $P$ to be precisely the zero pattern of $A$. In the following, we denote by $b_i$ the $i$-th row of a given matrix $B$, and by $NZ(B)$, the set of pairs $(i,j), 1 \leq i, j \leq n$ such that $b_{i,j} \neq 0$.

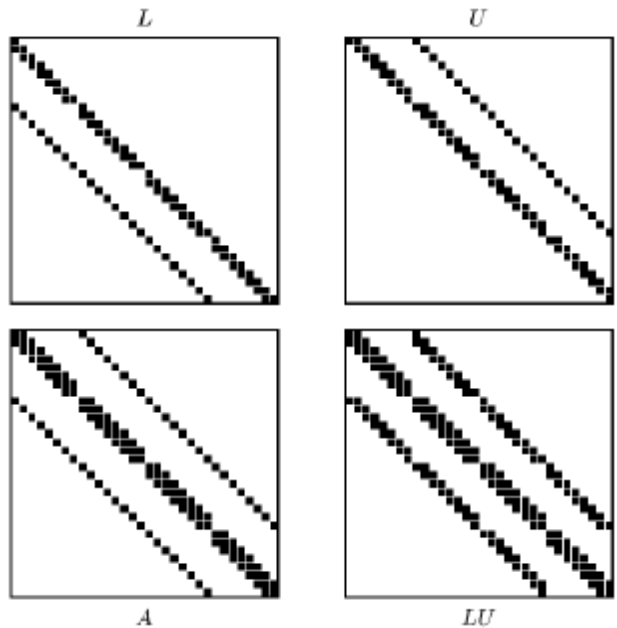

Fig 5.1: The ILU (0) factorization for a five-point matrix

The accuracy of the ILU(0) incomplete factorization may be insufficient to yield an adequate rate of convergence. More accurate Incomplete LU factorizations are often more efficient as well as more reliable. These more accurate factorizations will differ from ILU(0) by allowing some fill-in. Thus, ILU(1) keeps the "first order fill-ins" a term which will be explained shortly. To illustrate ILU($p$) with the same example as before, the ILU(1) factorization results from taking $P$ to be the zero pattern of the product $LU$ of the factors $L,U$ obtained from ILU(0). This pattern is shown at the bottom right of Figure 5.1. Pretend that the original matrix has this "augmented" pattern $NZ_1(A)$. In other words, the fill-in positions created in this product belong to the augmented pattern $NZ_1(A)$, but their actual values are zero. The new pattern of the matrix $A$ is shown at the bottom left part of Figure 5.2. The factors $L_1$ and $U_1$ of the ILU (1) factorization are obtained by performing an ILU (0) factorization on this "augmented pattern" matrix. The patterns of $L_1$ and $U_1$ are illustrated at the top of Figure 5.2. The new LU matrix shown at the bottom right of the figure has now two additional diagonals in the lower and upper parts.

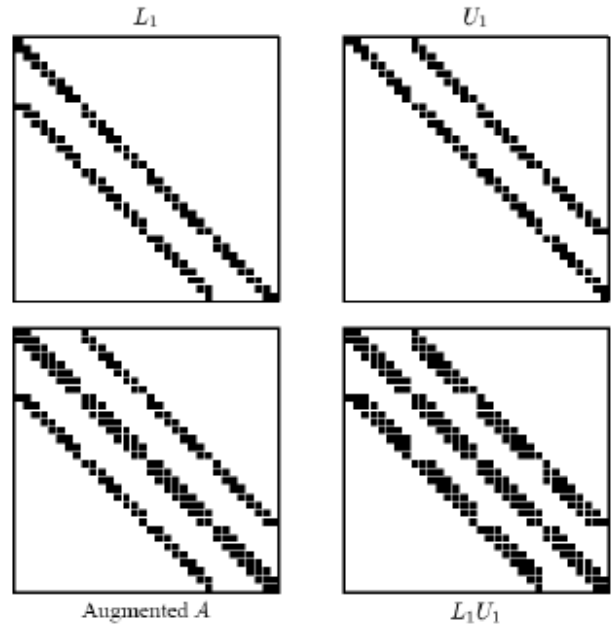

Fig 5.2: The ILU(1) factorization

**Algorithm 5.1: ILU (p)**
1. For all nonzero elements $a_{ij}$ define lev ($a_{ij}$) =0
2. For i = 2,...,n Do:
3. For each k =1,...,i −1 and for lev ($a_{ij}$) ≤ p Do:
4. Compute $a_{ik}=a_{ik}/a_{kk}$
5. Compute : $a_{i*} := a_{i*} - a_{ik}a_{k*}$
6. Update the level of fill of the nonzero $a_{i,j}$
7. EndDo
8. Replace any element in row i with lev ($a_{ij}$) > p by zero
9. EndDo

## 5.2: Krylov Subspace Method

**Table 5.2:** Different Krylov Subspace Algorithm:

| Krylov Sybspace Method | KSPType |
|---|---|
| Richardson | KSPRICHARDSON |
| Chebychev | KSPCHEVBYCHEV |
| Conjugate Gradients | KSPCG |
| GMRES | KSPGMRES |
| Bi-CG-stab | KSPBCGS |
| Transpose-free Quasi Minimal-Residual | KSPTFQMR |
| Conjugate Residuals | KSPCR |
| Conjugate Gradient Squared | KSPCGS |
| Bi-Conjugate Gradient | KSPBICG |
| Minimum Residual Method | KSPMINRES |
| Flexible GMRES | KSPFGMRES |
| Least Squares Method | KSPLSQR |
| SYMMLQ | KSPSYMMLQ |
| LGMRES | KSPLGMRES |
| Conjugate gradient on the normal equations | KSPCGNE |

**BiCGStab**

The Bi-Conjugate Gradient Stabilized (BiCGStab) algorithm is a variation of Conjugate Gradient Squared (CGS). As CGS is based on squaring the residual polynomial, and, in cases of irregular convergence, this may lead to substantial build-up of rounding errors, or possibly even overflow. BICGSTAB was developed to remedy this difficulty.\

**The Algorithm:**

Choose $x^{(0)}$
$r^{(0)} = b - Ax^{(0)}$
Choose $\hat{r}$ (usually $\hat{r} = r^{(0)}$)
for $i = 1, 2, \ldots$
    $\rho_{i-1} = \hat{r}^T r^{(i-1)}$
    if $\rho_{i-1} = 0$ method fails
    if $i = 1$
        $p^{(1)} = r^{(0)}$
    else
        $\beta_{i-1} = \frac{\rho_{i-1}}{\rho_{i-2}} \frac{\alpha_{i-1}}{\omega_{i-1}}$
        $p^{(i)} = r^{(i-1)} + \beta_{i-1}(p^{(i-1)} - \omega_{i-1} v^{(i-1)})$
    endif
    Solve $M\hat{p} = p^{(i)}$ ; $v^{(i)} = A\hat{p}$
    $\alpha_i = \frac{\rho_{i-1}}{\hat{r}^T v^{(i)}}$
    $s = r^{(i-1)} - \alpha_i v^{(i)}$
    if $\|s\|$ is small enough then
        $x^{(i)} = x^{(i-1)} + \alpha_i \hat{p}$ and stop
    Solve $Mz = s$ ; $t = Az$
    $\omega_i = \frac{s^T t}{t^T t}$
    $x^{(i)} = x^{(i-1)} + \alpha_i \hat{p} + \omega_i z$
    Check for Convergence
    if $\omega_i = 0$ stop
    $r^{(i)} = s - \omega_i t$
end

# 6. Results and Discussion:

## 6.1: Solution Profile and the Error Plot:

| Parameter | $\varepsilon$ | X | Y | Tolerance | No. of Mesh Size on Both Direction |
|---|---|---|---|---|---|
| Magnitude | 4 | 4 | 4 | 1e-20 | 20 |

**Type of Solution:** ILU Preconditioned BiCGSTAB Method

**Type of ILU used:** ILU (0) : Ref. *Iterative Methods for Sparse Linear Systems* by Y. Saad

| | Dirichlet Boundary Condition |
|---|---|
| **Solution P**<br><br>Upwind (Hybrid) Scheme | 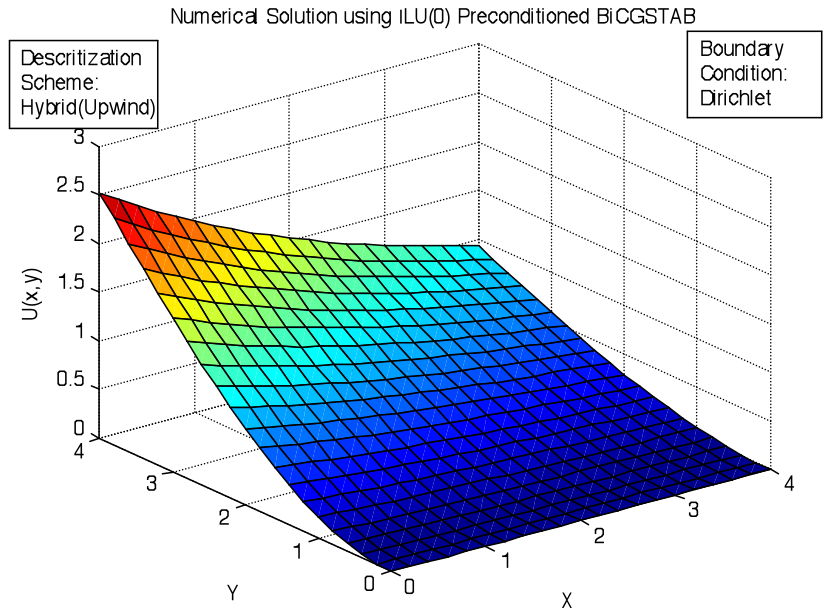 |
| Centered Scheme | 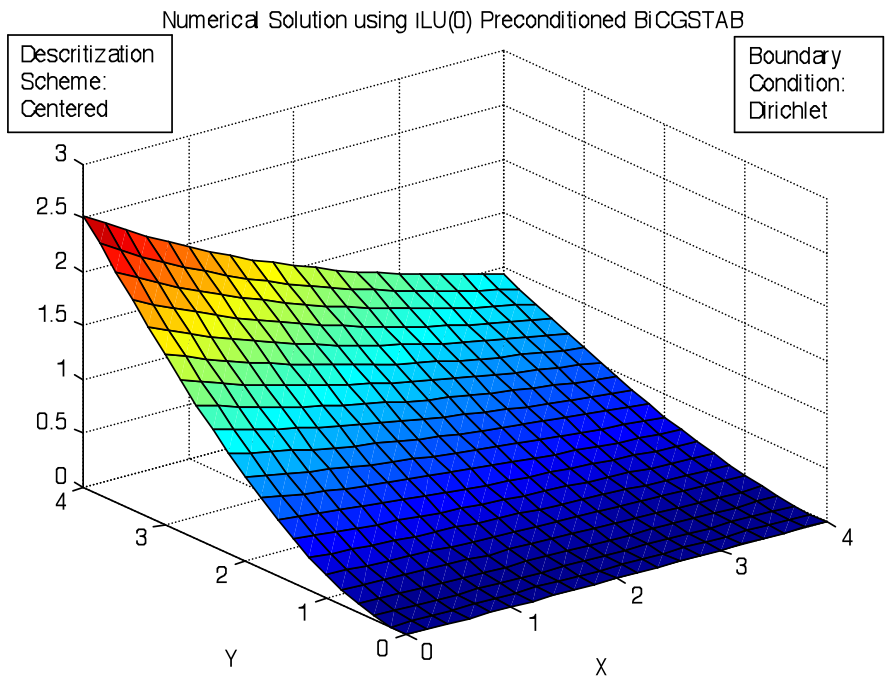 |

**Error Plot:**

| | Dirichlet Boundary Condition |
|---|---|
| | |

| Upwind (Hybrid) Scheme | 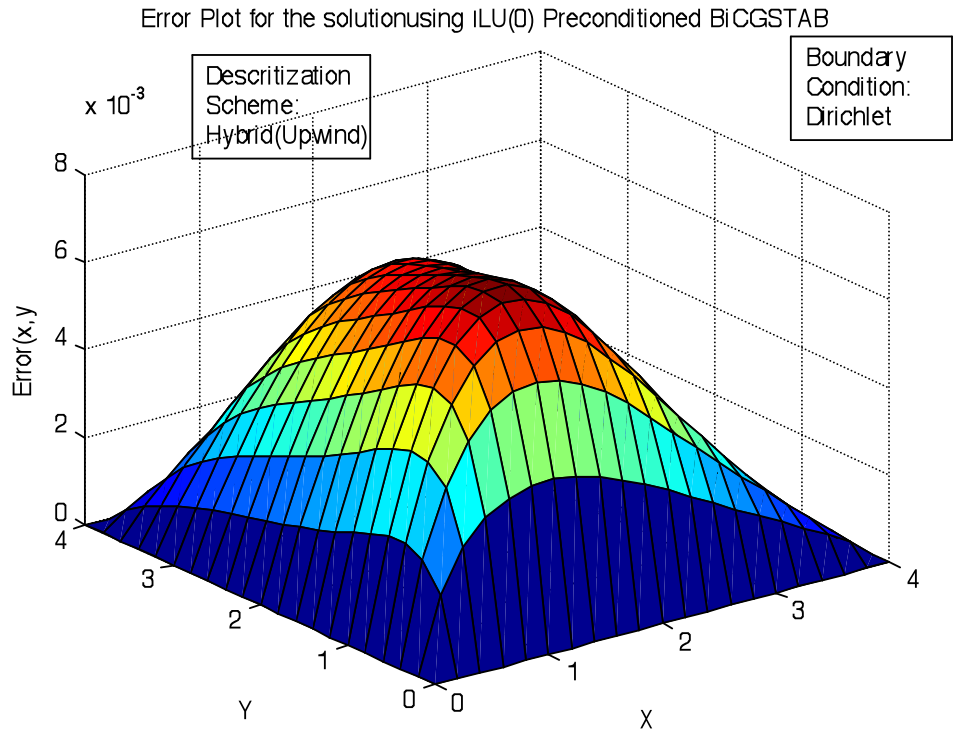 |
|---|---|
| Centered Scheme | 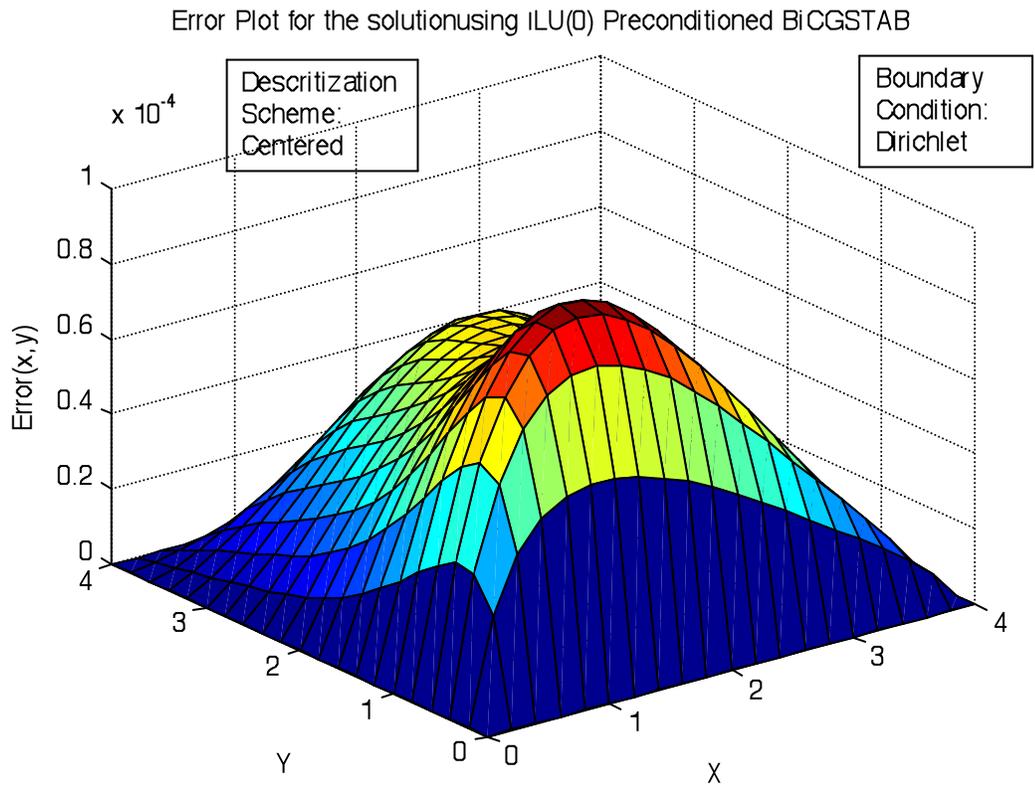 |

**Solution Plot:**

| | Neumann Boundary Condition |
|---|---|
| Upwind (Hybrid) Scheme | 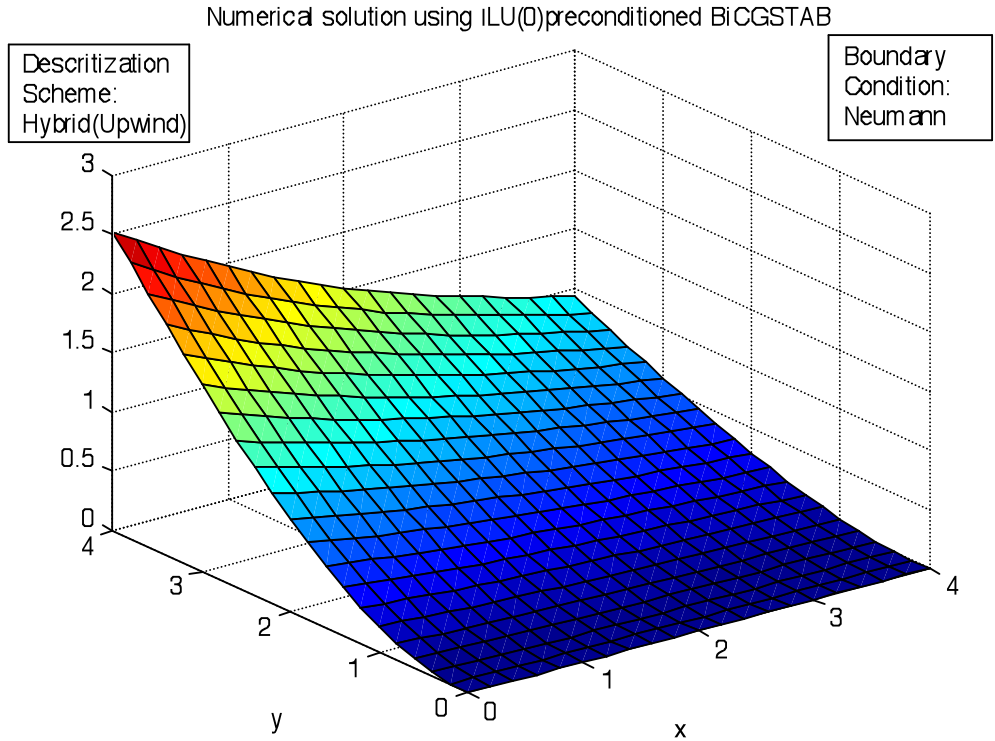 |
| Centered Scheme (for the governing eqn.) | 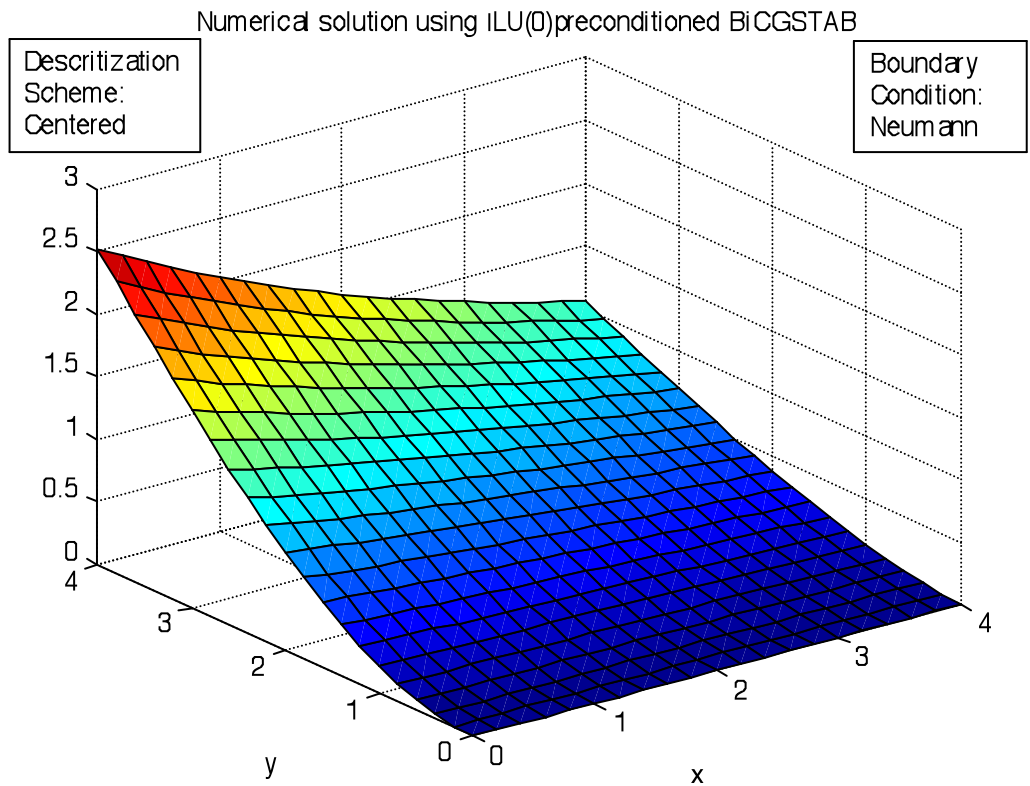 |

**Error Plot:**

| | Neumann Boundary Condition |
|---|---|
| Upwind (Hybrid) Scheme | 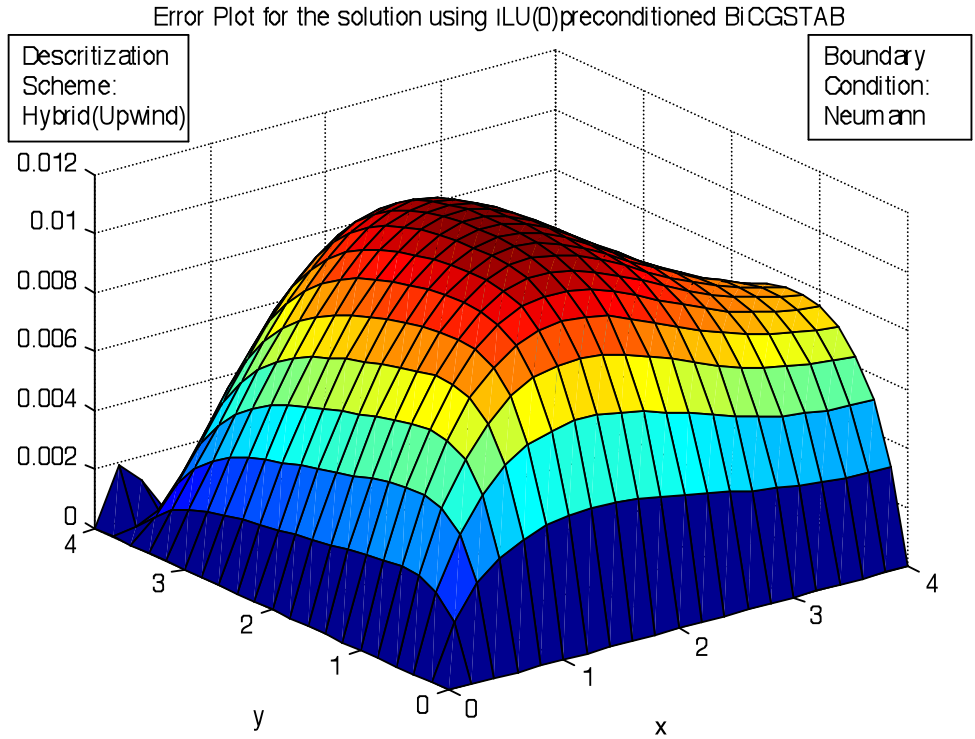 |
| Centered Scheme (for the governing eqn.) | 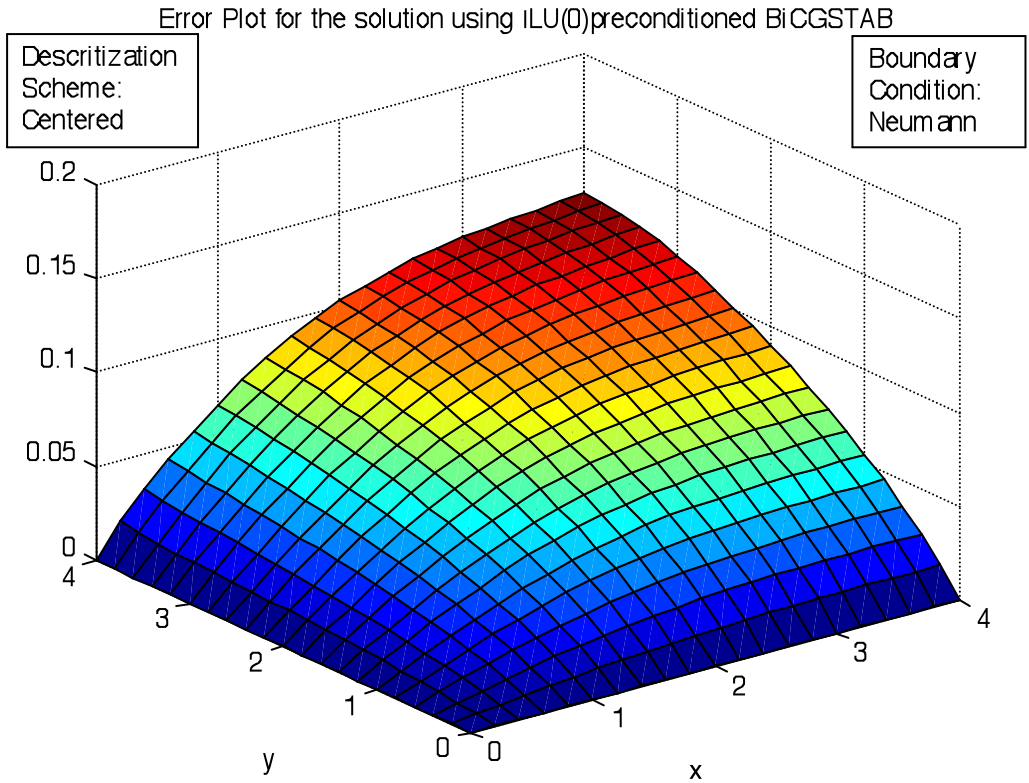 |

**Comment on the Result:**

*Brief Statistics:*

|  | Dirichlet Boundary Condition | | Neumann Boundary Condition | |
|---|---|---|---|---|
|  | Hybrid Scheme | Centered Scheme | Hybrid Scheme | Centered Scheme ( for the governing eqn.) |
| No. of Iteration Required | 24 | 24 | 30 | 36 |
| Maximum Error | 0.0073 | 9.51*1e-05 | 0.0110 | 0.1256 |

➢ For low value of epsilon, the Neumann condition does not work. The matrix may be become badly conditioned. For the present case the author has observed that the for epsilon < 2, the Neumann condition does not work.

➢ As mentioned in the above table, iteration required for the Neumann case is more as compared to its Dirichlet counterpart.

➢ For the Dirichlet boundary condition, the number of iteration for the Hybrid and Centered scheme are same. However, the maximum error for the Centered scheme is less. It can be judged using the figure bellow.

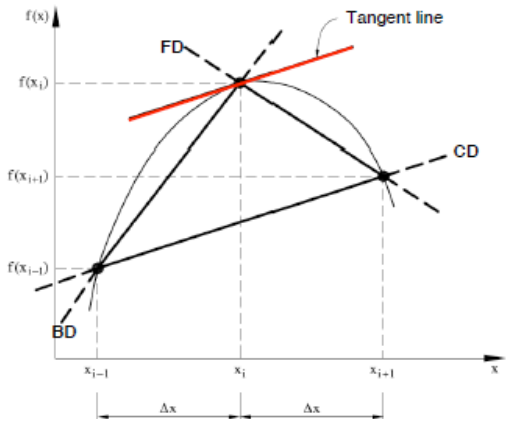

It can be observed that the tangent is better approximated using the Centered scheme as compared to the Forward or Backward scheme.

➢ With larger value of epsilon, no of iteration required for Neumann and Dirichlet is much different.

➢ With less number of less, the solution may diverge in case of Neumann condition (for epsilon very small).

➢ In case of Neumann condition, the error at the corner node which does not appear in the problem formulation will depend on the methodology of computing the corner nodes for the condition of hybrid scheme. When the corner node is approximated with the average of the value obtained from the adjacent two nodes, then the error is different from the case when it is obtained for the case of obtaining the value from one adjacent node.

| 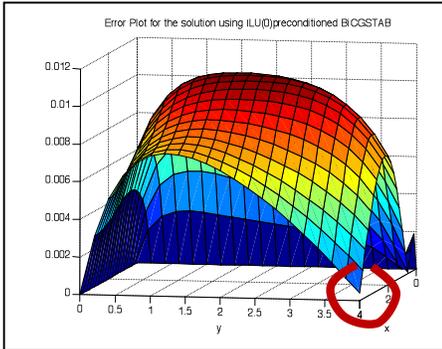 | 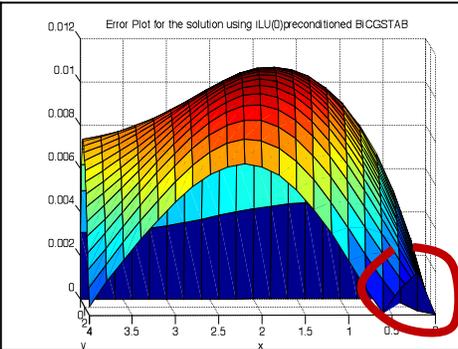 | 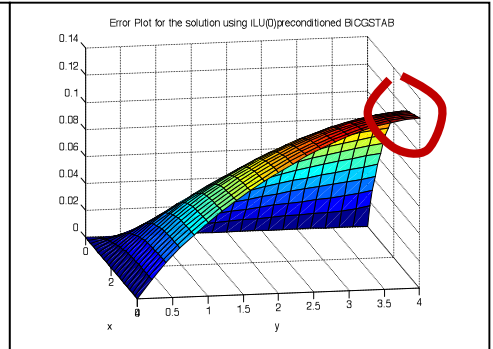 |
|---|---|---|
| Hybrid: Corner node obtained from the avg. of the value obtained from node known adjacent nodes. | Hybrid: Corner node obtained from the value obtained from one adjacent known node. | Centered (for the governing equation): Error at the corner node not affected considering case of obtaining value from two nodes or one node |

## 6.2: Variation of Response w.r.t. ε:

| Parameter | X | Y | Tolerance | No. of Mesh Size on Both Direction |
|---|---|---|---|---|
| **Magnitude** | 4 | 4 | 1e-20 | 20 |

**Type of Solution:** ILU Preconditioned BiCGSTAB Method

**Type of ILU used:** ILU (0) : Ref. *Iterative Methods for Sparse Linear Systems* by *Y. Saad*

**Boundary Condition:** Dirichlet

| ε | Hybrid Scheme | Centered Scheme |
|---|---|---|

| ε | Hybrid Scheme | Centered Scheme |
|---|---|---|
| 0.0001 | 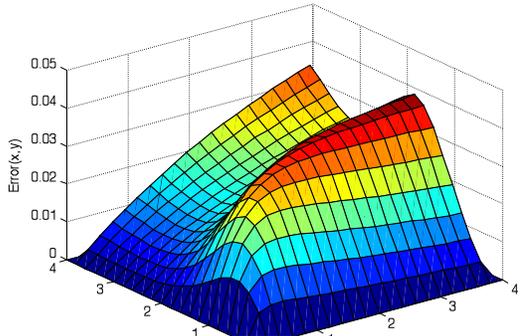 | SOLUTION OSCILLATES AND DOES NOT CONVERGE EVEN AFTER 1000 ITERATION |
| 0.001 | 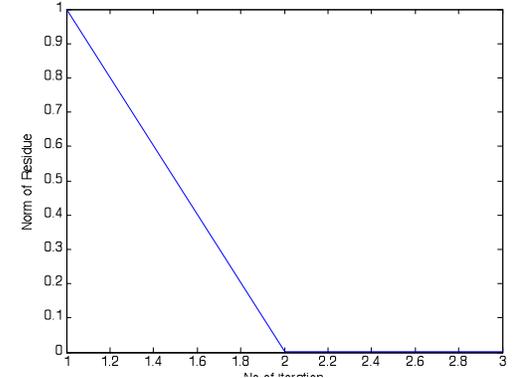 | SOLUTION OSCILLATES AND DOES NOT CONVERGE EVEN AFTER 1000 ITERATION |
| ε | Hybrid Scheme | Centered Scheme |

| ε | Hybrid Scheme | Centered Scheme |
|---|---|---|
| 0.01 | 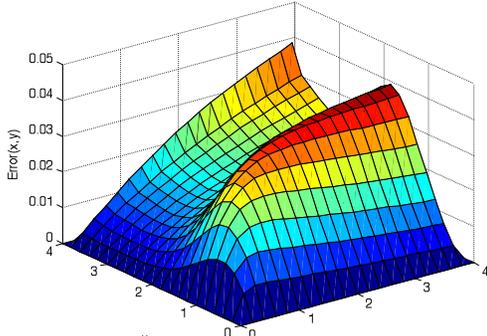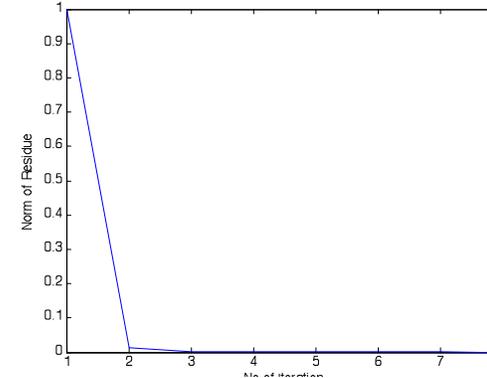 | 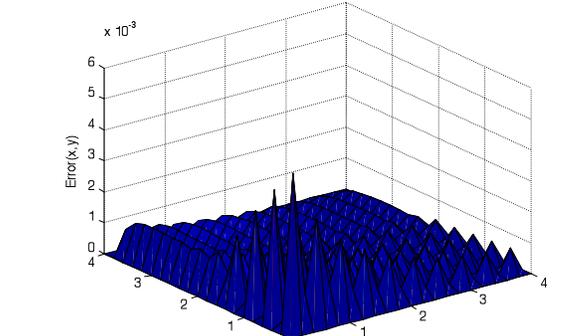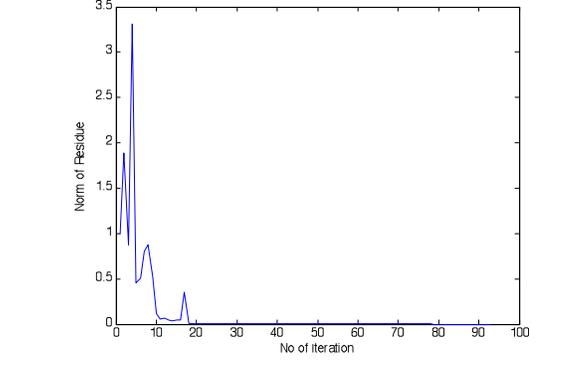 |
| 0.05 | 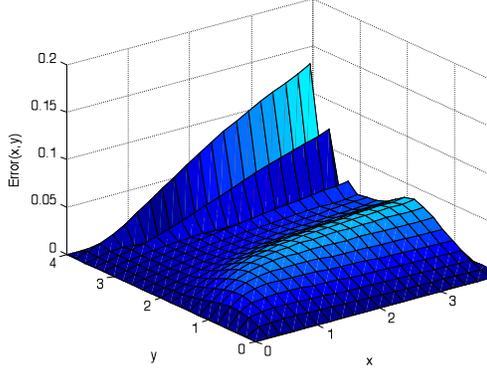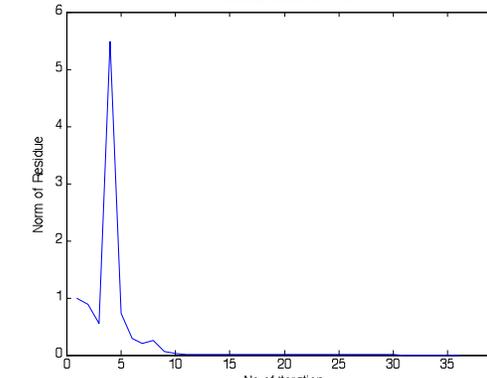 | 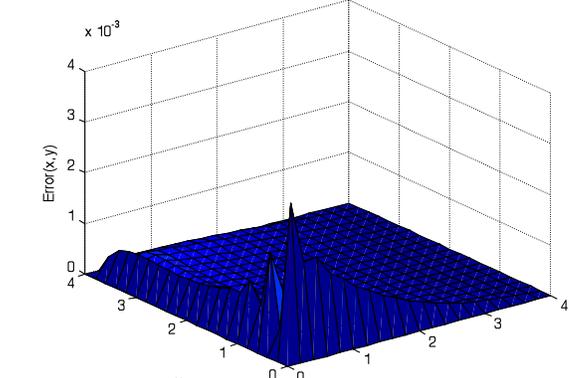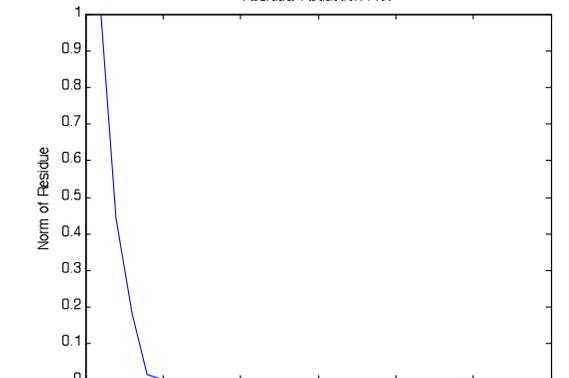 |

| ε | Hybrid Scheme | Centered Scheme |
|---|---|---|
| 1 | 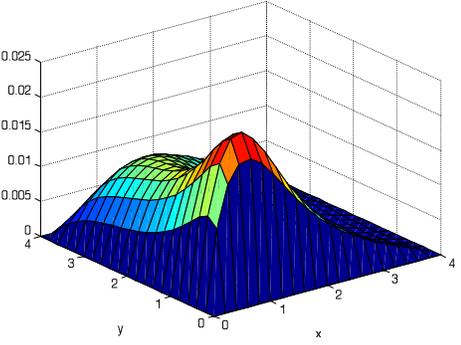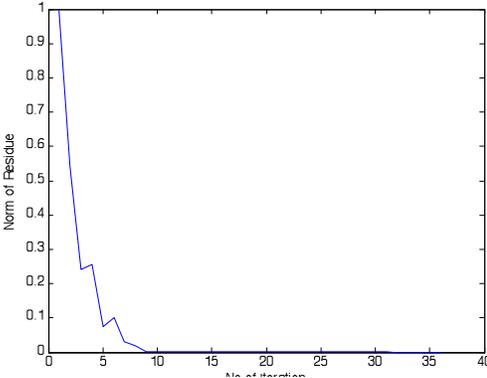 | 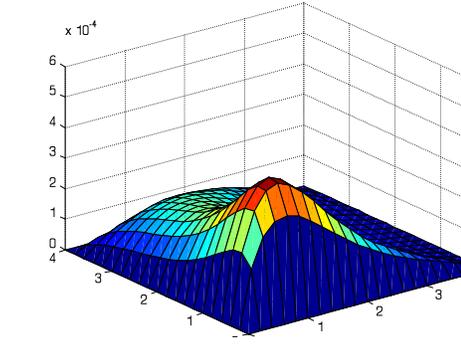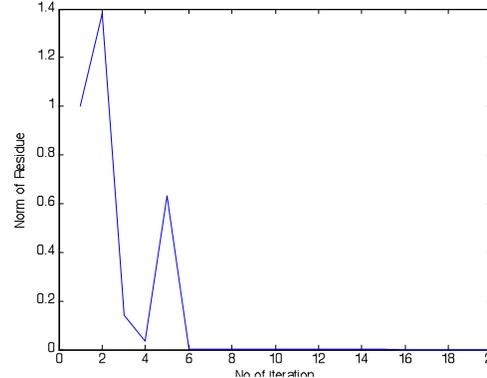 |
| 5 | 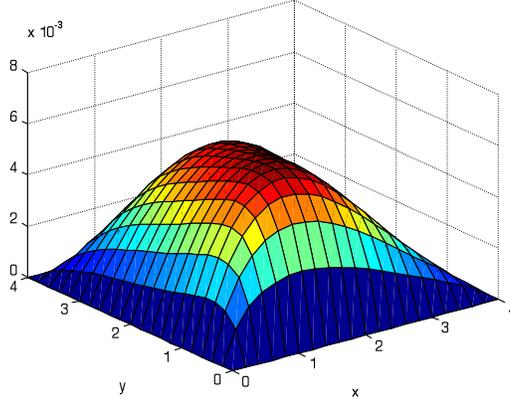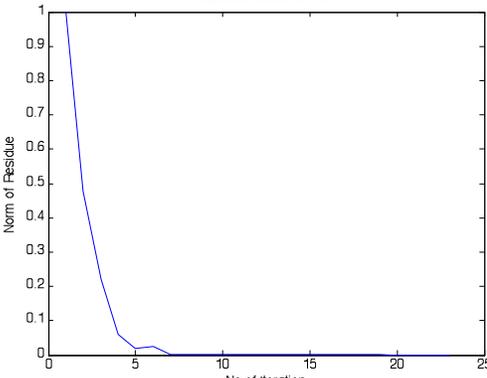 | 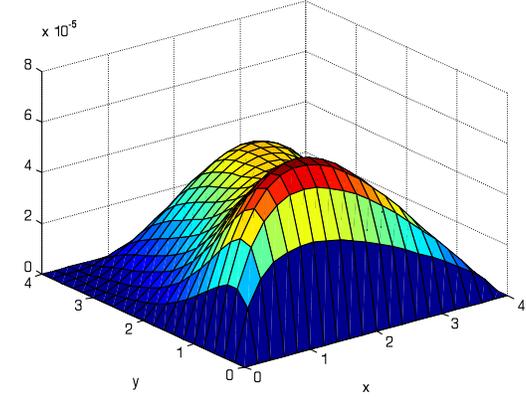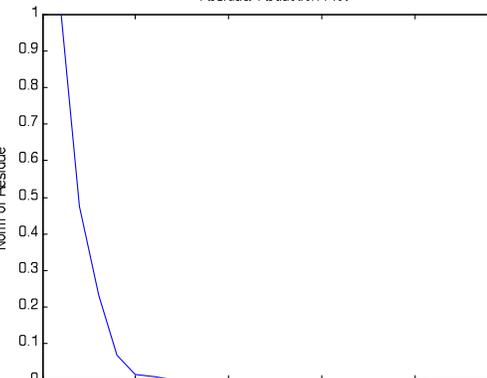 |
| 10 | 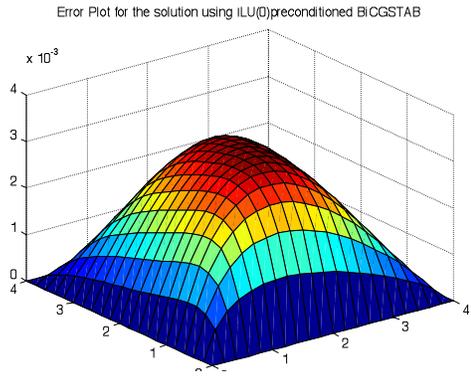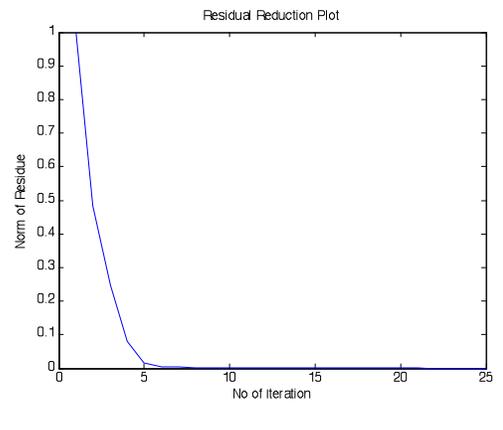 | 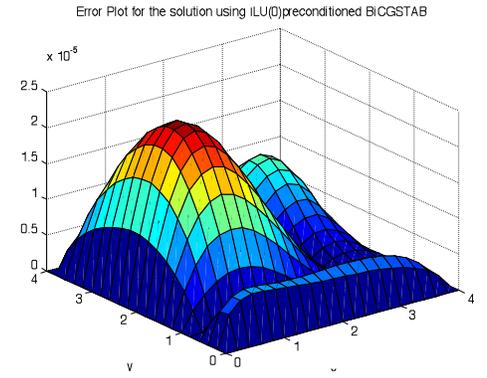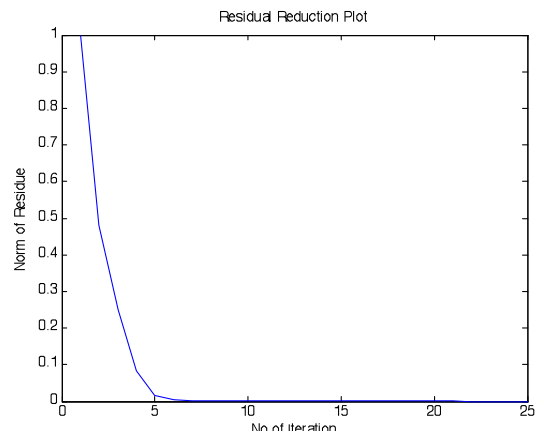 |
| 50 | 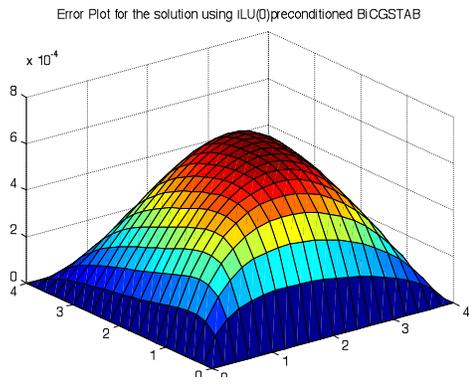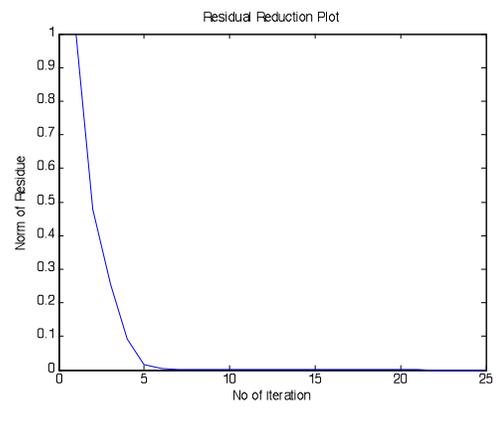 | 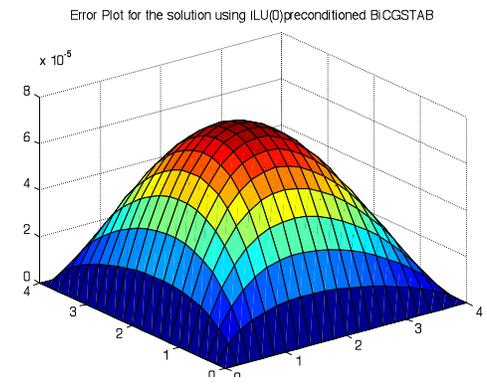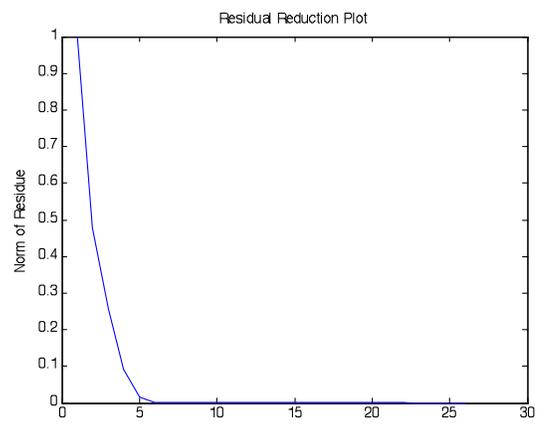 |

**Comment on the Results:**

*Brief Statistics (NC: Not Converged)*

| $\varepsilon$ | | 0.0001 | 0.001 | 0.01 | 0.05 | 1 | 5 | 10 | 50 |
|---|---|---|---|---|---|---|---|---|---|
| Iteration Required | Hybrid Scheme | 4 | 5 | 8 | 37 | 37 | 24 | 26 | 26 |
| | Centered Scheme | NC | NC | 94 | 30 | 21 | 24 | 26 | 27 |
| Maximum Error | Hybrid Scheme | 0.0406 | 0.0407 | 0.0415 | 0.1391 | 0.0234 | 0.0060 | 0.0032 | 6.2941e-04 |
| | Centered Scheme | NC | NC | 0.0051 | 0.0031 | 4.2852e-04 | 6.4971e-05 | 2.3639e-05 | 7.7753e-05 |

> ➤ For the large value of ε, the convergence of the solution is much more ensured. The error is less for large ε. The stiffness matrix is well conditioned for large values of ε. The stiffness matrix is much more diagonally dominant for the case of higher values of epsilon.

> ➤ For very low values of ε (0.0001-0.001), the centered scheme does not converged. The hybrid scheme converged with very less iteration. For the case of ε=0.01, the centered scheme converged but with large number of iteration and with high oscillation before convergence.

> ➤ For the case of ε=0.5, the hybrid scheme does not converge. However, the centered scheme converged.

> ➤ For higher value of ε, both the scheme converges. However, the error in case of centered scheme is much less as compared to the hybrid scheme. The probable reason is same as mentioned in the previous section.

> ➤ For the low value of ε, the oscillation in the convergence of the solution for the centered scheme can be described by the mesh Reynolds number. When the mesh Reynolds number > 2, the solution will have oscillation in the convergence.

## 6.3: Mesh Reynolds Number:

| Parameter | X | Y | Tolerance | ε | No. of Mesh Size on Both Direction |
|---|---|---|---|---|---|
| Magnitude | 4 | 4 | 1e-20 | 1 | 20 |

**Type of Solution:** ILU Preconditioned BiCGSTAB Method

**Type of ILU used:** ILU (0) : Ref. *Iterative Methods for Sparse Linear Systems* by *Y. Saad*

**Boundary Condition:** Dirichlet

**Discritization Scheme:** Hybrid

| ah/ε | bh/ε |
|---|---|

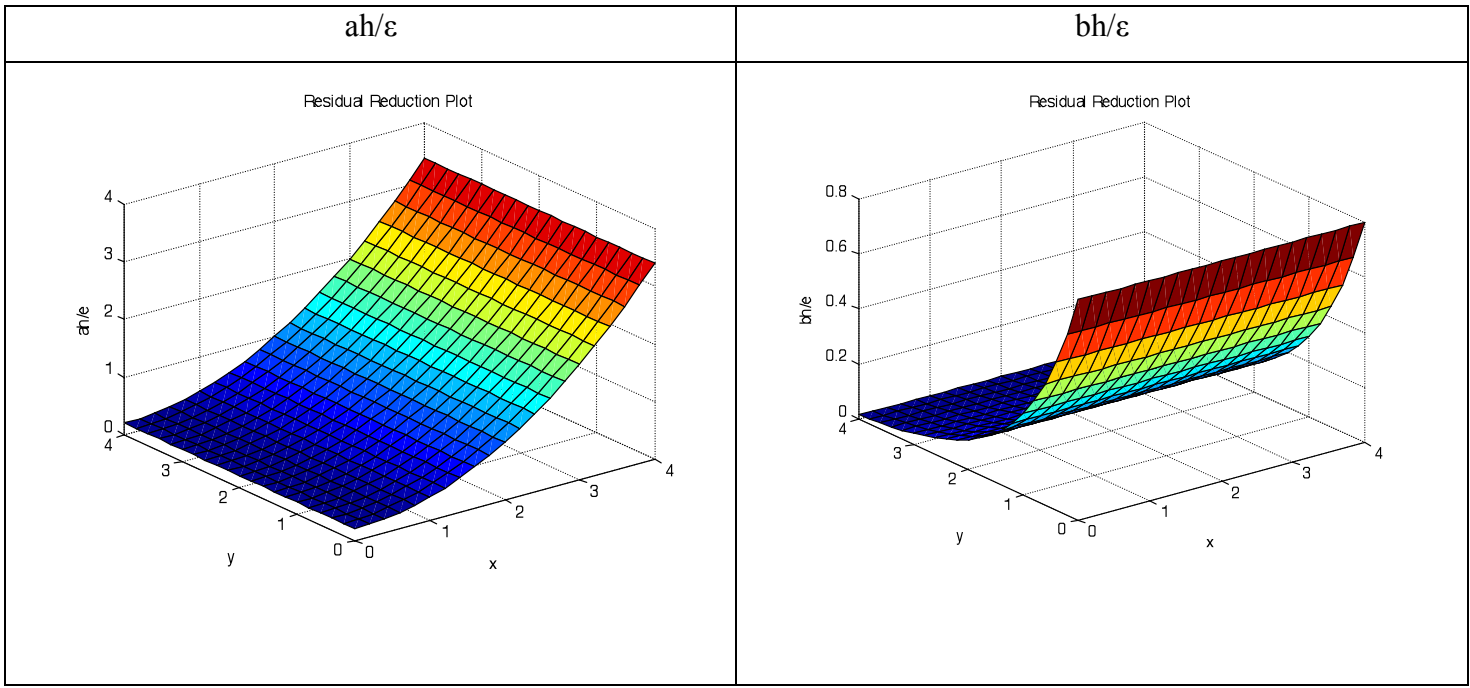

**Comment:**

- The mesh Reynolds number does not depend on the descritization scheme or problem type. It depends on the epsilon and the mesh size only.

- The mesh Reynolds number is associated with the stability of the solution. For the centered scheme for having a stable solution, the mesh Reynolds number should be <2.

- Hence for a given value of ε, the stability of the solution can be obtained with decrease in the mesh size.

- For mesh Reynolds number >2, the scheme is not monotonic. It has 2Δx oscillation-**nequest oscillation.**(Oscillation at the smallest wave length)

## 6.4: Residual Reduction Plot:

| Parameter | X | Y | Tolerance | $\varepsilon$ | No. of Mesh Size on Both Direction |
|---|---|---|---|---|---|
| Magnitude | 4 | 4 | 1e-20 | 4 | 20 |

**Discritization Scheme:** Hybrid

| Solution Parameter | Dirichlet | Neumann |
|---|---|---|
| ILU (Direct) (Ref. Prof JP) No Acceleration | Residual Reduction Plot (residue decays from 1 to 0 over ~250 iterations) | Residual Reduction Plot (residue decays from 1 to 0 over ~2000 iterations) |
| ILU (Ref. Prof Y. Saad) No Acceleration | Residual Reduction Plot, $\times 10^{304}$, spike to ~16 near iteration 300 | Residual Reduction Plot, $\times 10^{303}$, spike near iteration 120 |
| ILU (Ref. Prof JP) Acceleration | Residual Reduction Plot, Norm of Residue peaks ~2.1 then decays by iteration 70 | Residual Reduction Plot, Norm of Residue peaks ~3.7 then decays by iteration 90 |

| Solution Parameter | Dirichlet | Neumann |
|---|---|---|
| ILU (Ref. Prof Y. Saad) Acceleration | 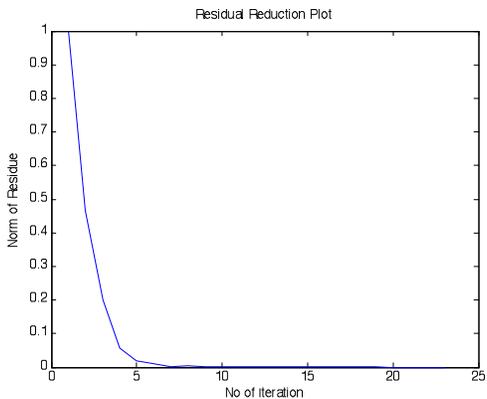 | 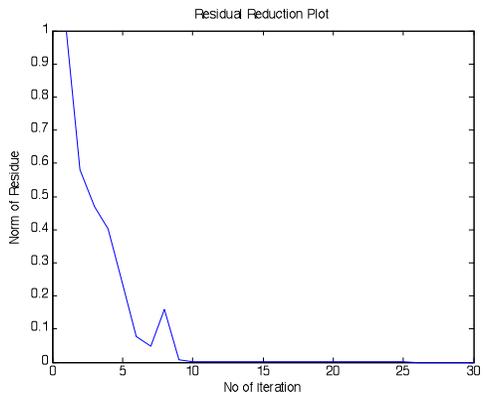 |
| No Preconditioned Acceleration | 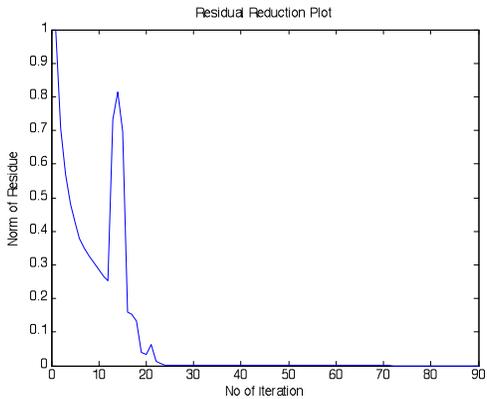 | 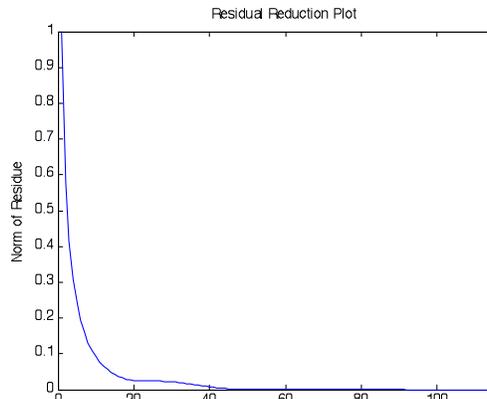 |

**6.5: Amplification Factor Plot:** (Flow parameter is same as before)

The amplification factor is computed as: $\dfrac{\|r^{k+1}\|}{\|r^k\|}$, Where the index k denotes the number of iteration. The amplification factor is expected to converge to a value less than 1.

| Solution Parameter | Dirichlet | Neumann |
|---|---|---|
| ILU (Direct) (Ref. Prof JP) No Acceleration | 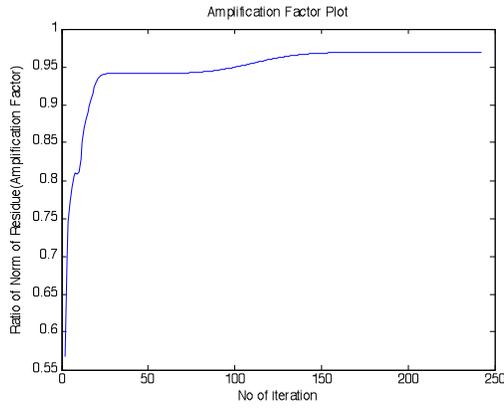 | 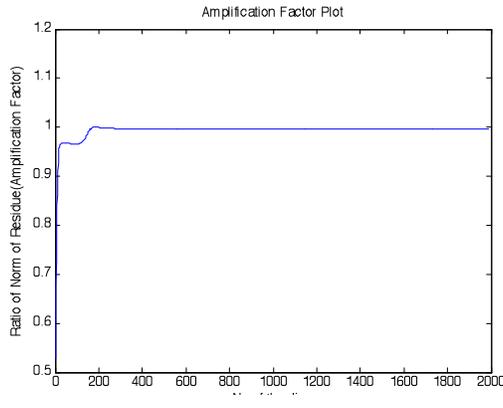 |
| ILU (Ref. Prof Y. Saad) No Acceleration | 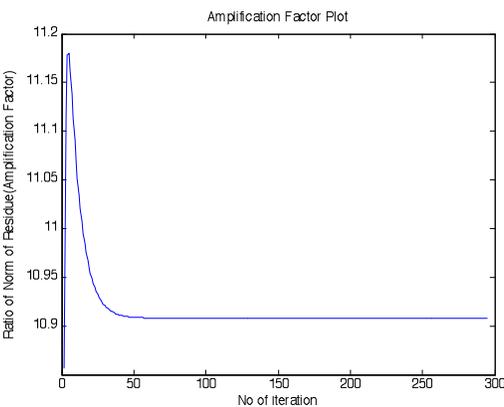 | 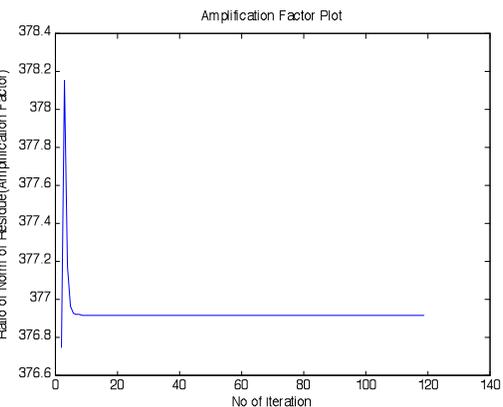 |
| ILU (Ref. Prof JP) Acceleration | 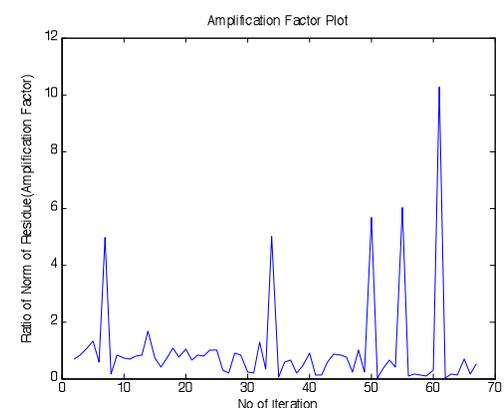 | 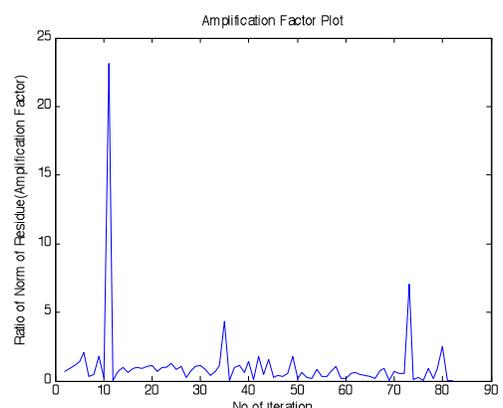 |

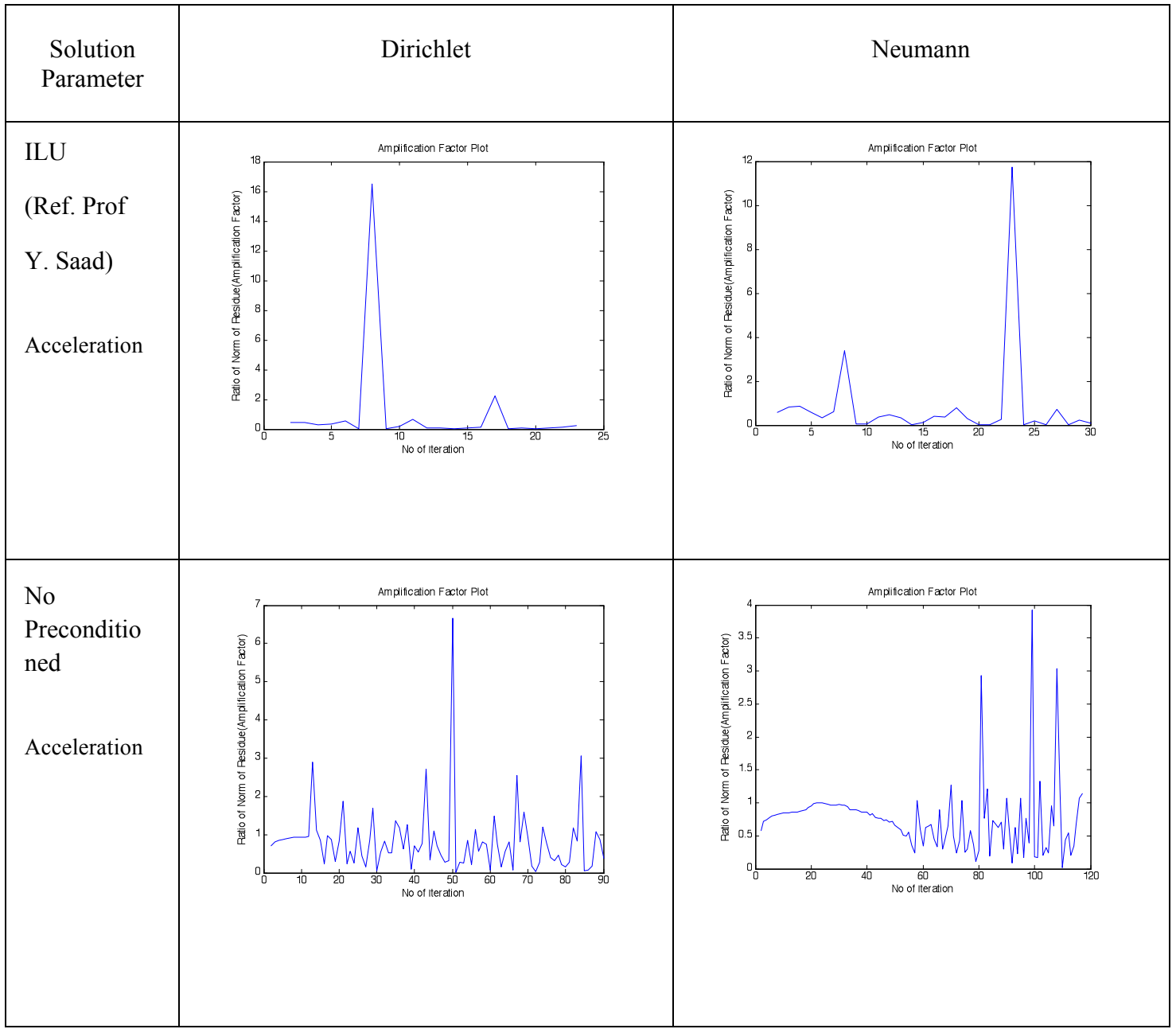

| Solution Parameter | Dirichlet | Neumann |
|---|---|---|
| ILU (Ref. Prof Y. Saad) Acceleration | | |
| No Preconditioned Acceleration | | |

Brief Statistics

| | | ILU(Direct) (Ref. Prof JP) No Acceleration | ILU (Ref.Prof Y.Saad) No Acceleration | ILU (Ref. Prof.JP) + BiCGSTAB | ILU (Ref. Prof.Y.Saad) +BiCGSATB | NO Preconditioned BiCGSTAB |
|---|---|---|---|---|---|---|
| No. of Iteration Required | Dirichlet | 207 | NC | 68 | 24 | 91 |
| | Neumann | 1989 | NC | 83 | 30 | 118 |
| Maximum Error | Dirichlet | 0.0072 | NC | 0.0073 | 0.0073 | 0.0073 |
| | Neumann | 0.0108 | NC | 0.0110 | 0.0110 | 0.0110 |

## 6.6: Error Plot for Different Methods: (Flow parameter is same as before)

| Solution Methods | Dirichlet | Neumann |
|---|---|---|
| ILU (Direct) (Ref. Prof JP) No Acceleration | 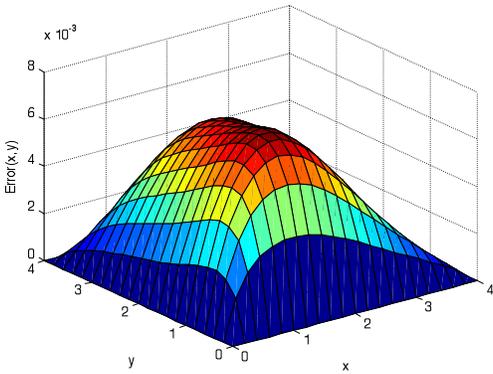 | 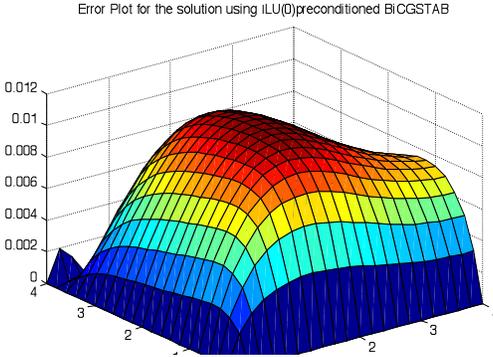 |
| ILU (Ref. Prof Y. Saad) No Acceleration | SOLUTION NOT CONVERGED | SOLUTION NOT CONVERGED |
| ILU (Ref. Prof JP) Acceleration | 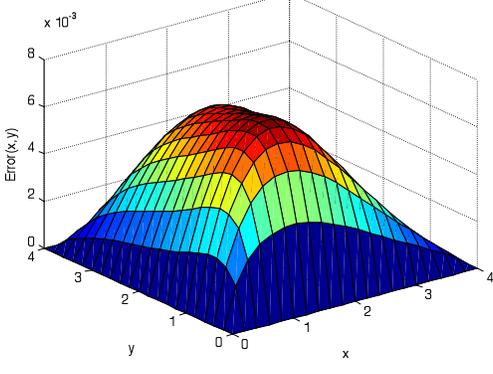 | 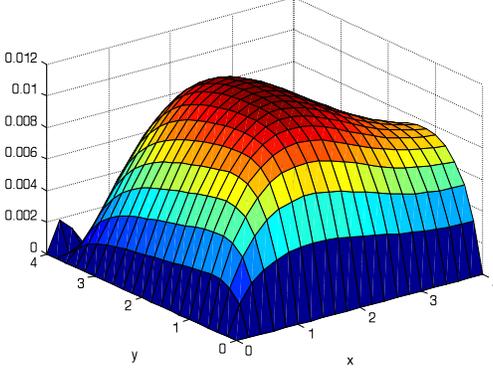 |

| Solution Parameter | Dirichlet | Neumann |
|---|---|---|
| ILU (Ref. Prof Y. Saad) Acceleration | 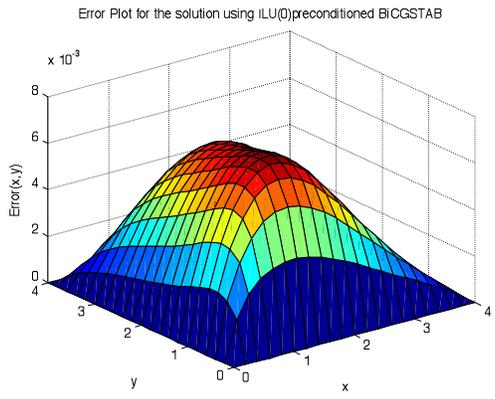 | 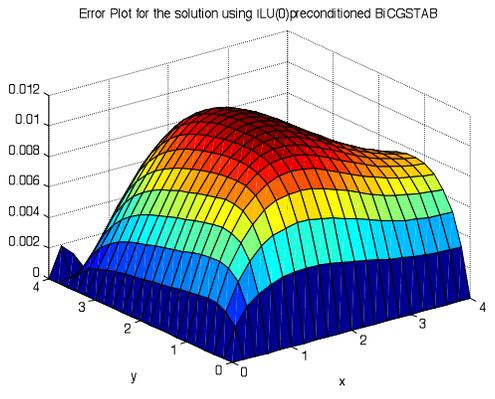 |
| No Preconditioned Acceleration | 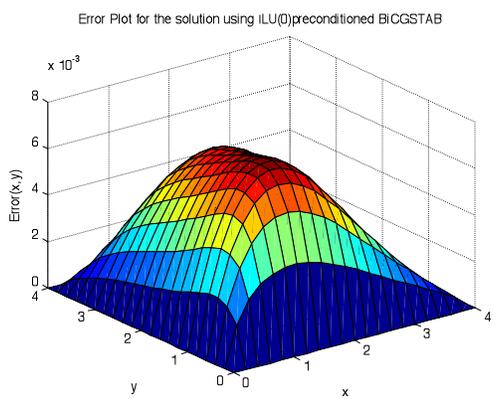 | 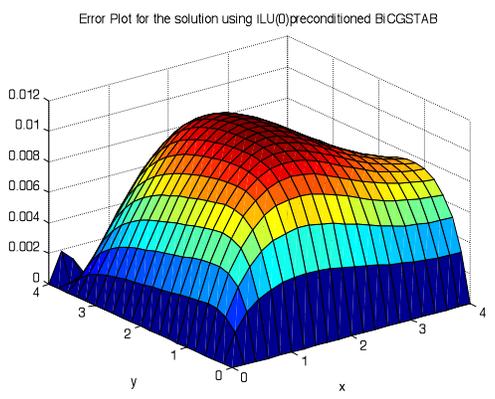 |

**Comment on the Result:**

➢ The maximum error for the different cases is almost same for both cases- Dirichlet and Neumann. The mode of convergence is considerably different for different methods of solution.

➢ For all the cases, the Neumann condition requires more number of iteration.

➢ The convergence is best in case of ILU(0) preconditioned BiCGSTAB.

➢ The ILU(0)+ NO BiCGSTAB has not converged. The reason is not known to the author.

## 6.7 Order of the Scheme:

| Parameter | X | Y | Tolerance | ε | No of cell in ech direction |
|---|---|---|---|---|---|
| Magnitude | 4 | 4 | 1e-20 | 4 | 20 |● 
| **Type of Solution:** ILU Preconditioned BiCGSTAB Method ||||||
| **Type of ILU used:** ILU (0) : Ref. *Iterative Methods for Sparse Linear Systems* by *Y. Saad* ||||||

|  | Dirichlet | Neumann (Upwind for Neumann Condition) |
|---|---|---|
|  | 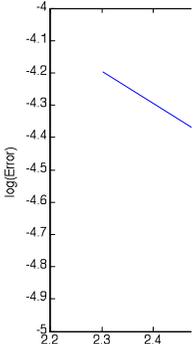 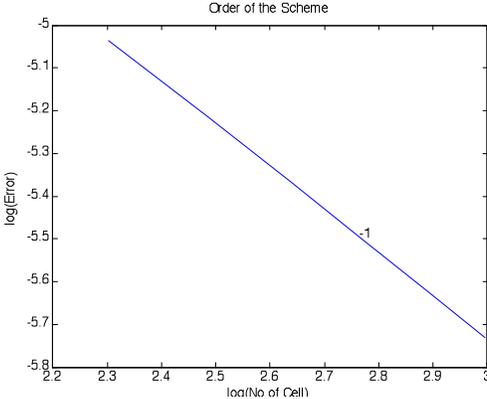 | 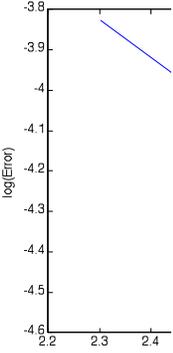 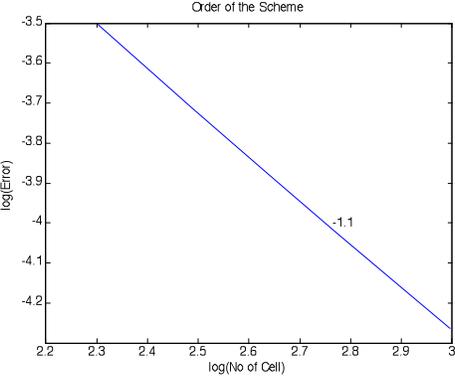 |
| Centered | 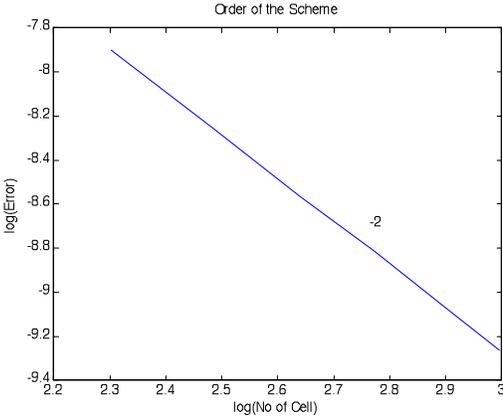 | 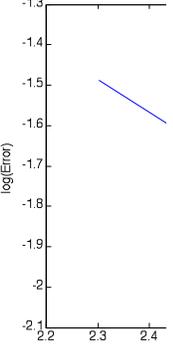 |

| | Dirichlet | Neumann (Upwind for Neumann Condition) |
|---|---|---|
| $\varepsilon=10$ Centered | 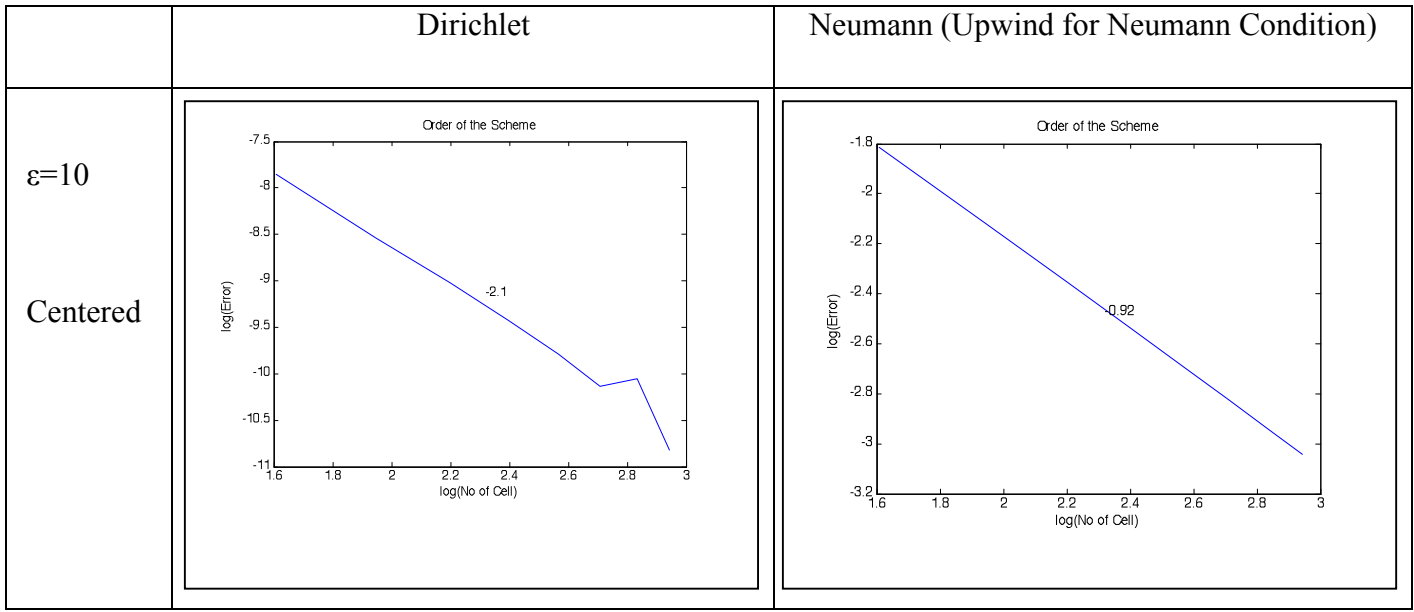 | |

**Comment on the Results:**

➢ Hybrid scheme has an order of around 1 and less then 2. The magnitude depends on the boundary condition. For particular boundary condition, it depends on the value of $\varepsilon$. Hybride scheme uses centered scheme for the second order derivative and upwind scheme for the first order derivative. Hence the order of the hybride descritization is expected in between 1 and 2.

➢ The centered scheme for the , in case of Dirichlet boundary condition, the order as expected is 2. The order has negligible perturbation with the change of $\varepsilon$.

➢ In case of Neumann condition, the centered scheme is used for the governing PDE. However, the Neumann boundary condition is approximated using the upwind scheme. The order of the system is around 1 and is influenced by the $\varepsilon$ parameter.

➢ The same comment is applicable for the Neumann condition with hybrid scheme for the governing PDE.

## 6.8 Response for Different Number of Cell:

| Parameter | X | Y | Tolerance | $\varepsilon$ | |
|---|---|---|---|---|---|
| **Magnitude** | 4 | 4 | 1e-20 | 4 | |

**Type of Solution:** ILU Preconditioned BiCGSTAB Method

**Type of ILU used:** ILU (0) : Ref. *Iterative Methods for Sparse Linear Systems* by *Y. Saad*

**Boundary Condition**: Dirichlet

| Mesh No. | Hybrid Scheme | Centered Scheme |
|---|---|---|
| 5 | 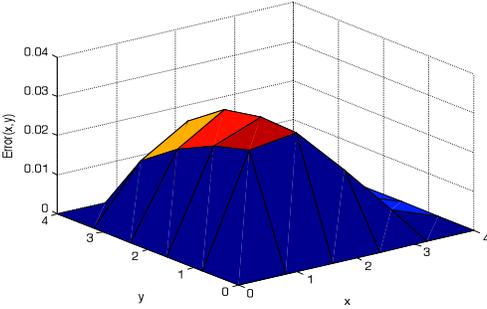<br>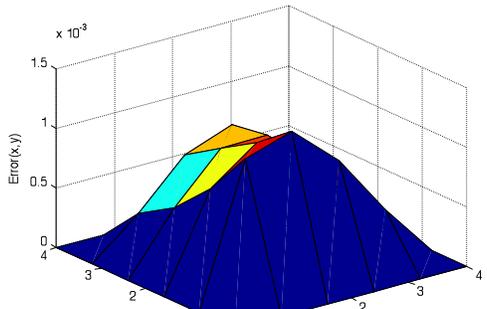 | 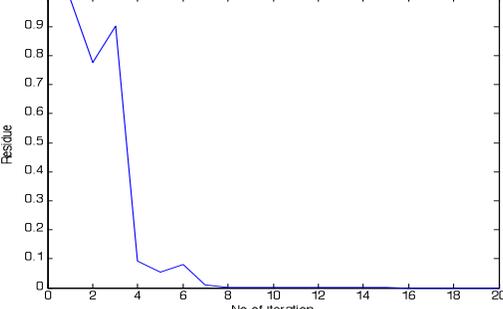<br>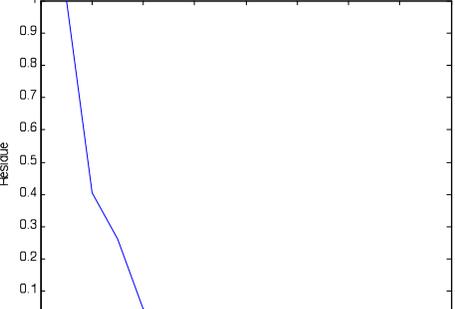 |
| 10 | 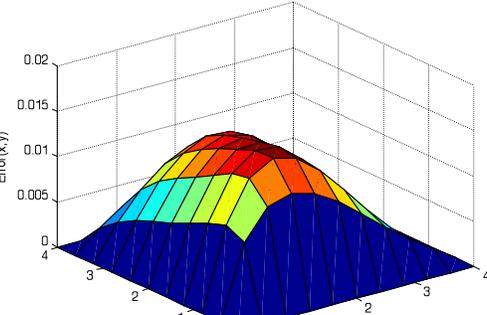<br>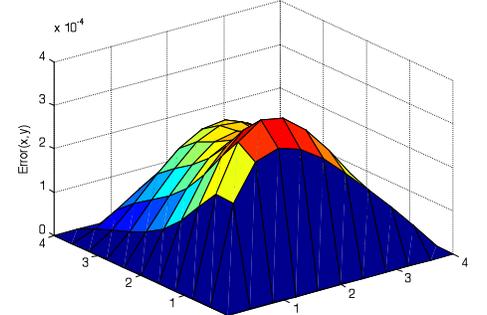 | 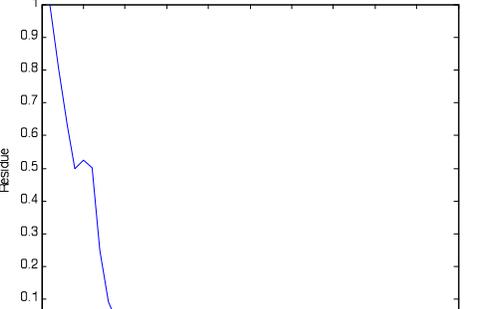<br>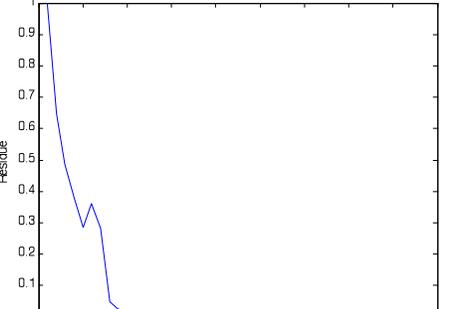 |

| Mesh No. | Hybrid Scheme | Centered Scheme |
|---|---|---|
| 15 | 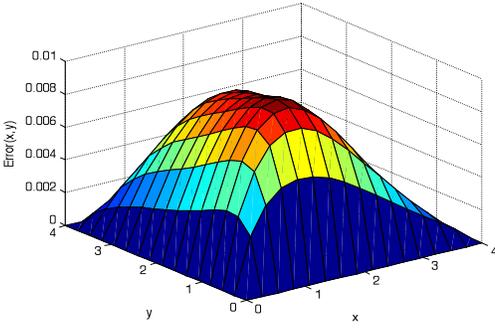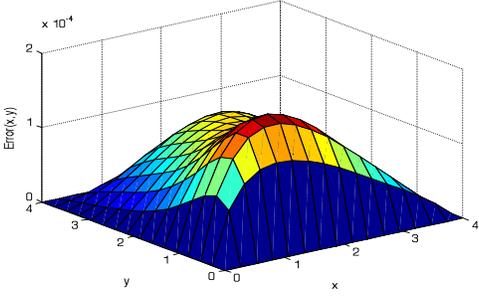 | 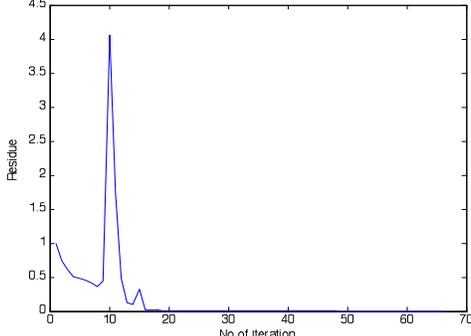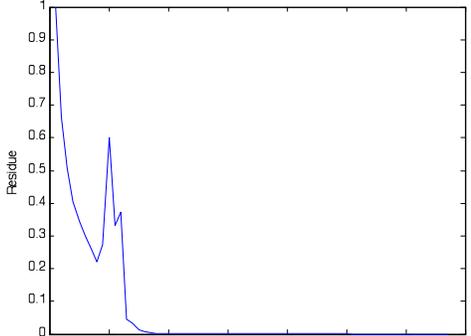 |
| 20 | 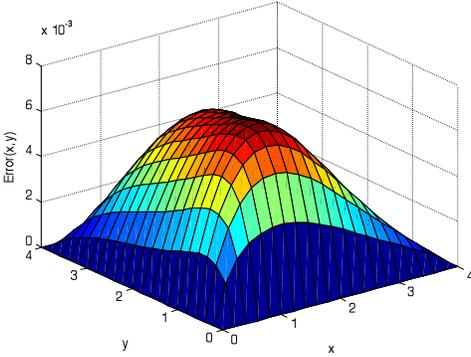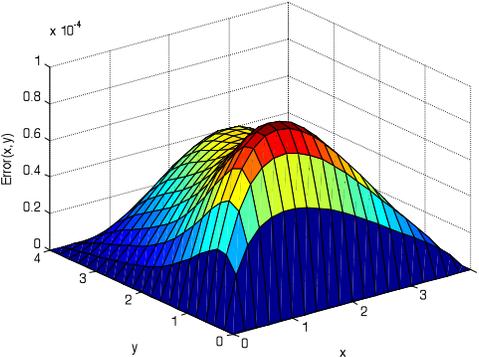 | 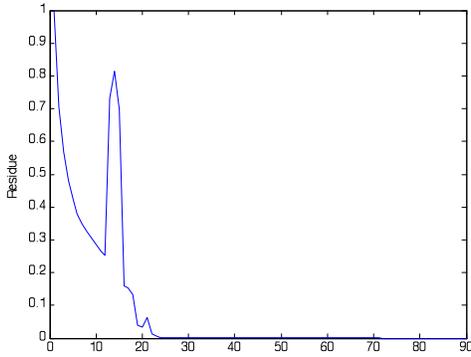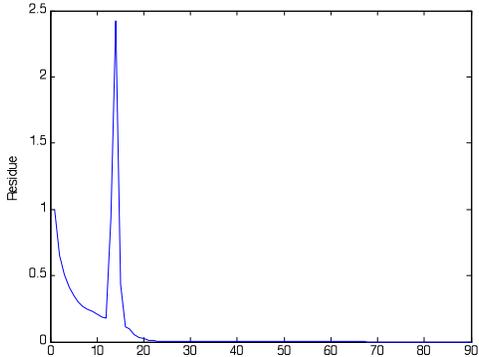 |

| Mesh No. | Hybrid Scheme | Centered Scheme |
|---|---|---|
| 25 | | |

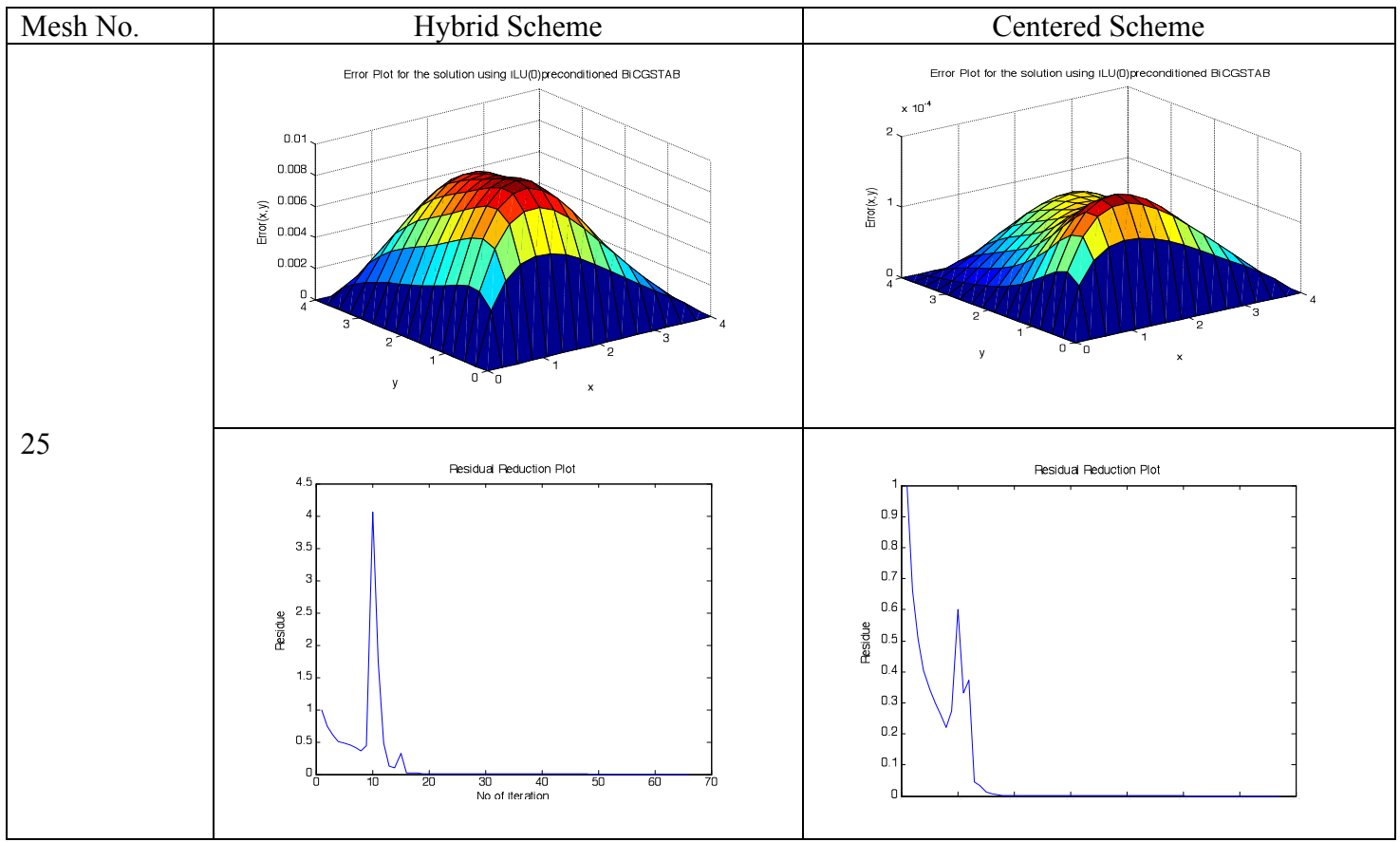

*Brief Statistics (NC: Not Converged)*

| No of Mesh | | 5 | 10 | 15 | 20 | 25 |
|---|---|---|---|---|---|---|
| **Iteration Required** | **Hybrid Scheme** | 21 | 51 | 66 | 91 | 111 |
| | **Centered Scheme** | 17 | 44 | 67 | 90 | 115 |
| **Maximum Error** | **Hybrid Scheme** | 0.0301 | 0.0150 | 0.0099 | 0.0073 | 0.0058 |
| | **Centered Scheme** | 0.0013 | 3.7061e-004 | 1.6819e-004 | 9.5110e-005 | 6.1190e-005 |
| | | | | | | |

**Comment on the Result:**

➢ The error gets reduced with the increase of cell number. However, the time of computation for the large number of cell is large.

➢ The error in each case for the centered case is less as compared to the hybrid case.

## 6.9: CPU Time required Plot:

| Parameter | X | Y | Tolerance | ε | No. of Mesh Size on Both Direction |
|---|---|---|---|---|---|
| Magnitude | 4 | 4 | 1e-5 | 4 | 20 |
| | | | | | |

**Comparison of Different Method of Solution:**

**Boundary Condition:** Dirichlet

**Discritization Scheme:** Hybrid

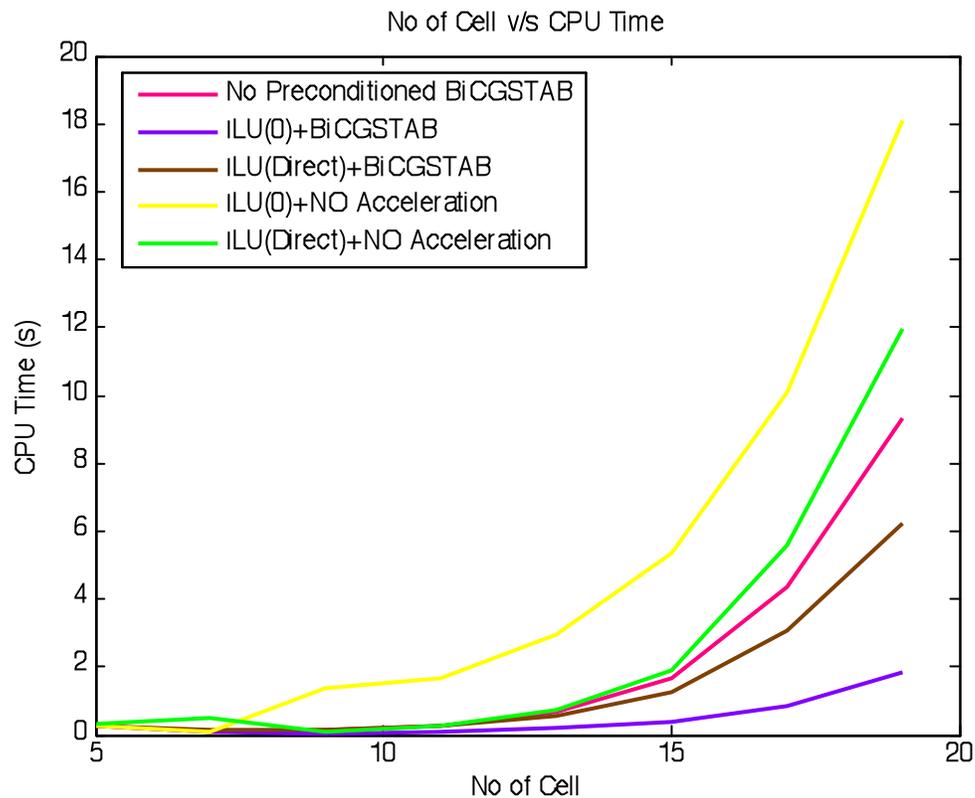

**Comment:**

➢ The time required for case of ILU(0)+BiCGSTAB is less as compared to all other cases.

➢ The maximum time required is for the case of ILU(0)+NO Preconditioned.

➢ Preconditioned BiCGSTAB takes less time as compared to the No preconditioned BiCGSTAB.

➢ Hence it can be concluded that Preconditioned acceleration method is the most suitable method.

**Comparison of Different Method of Discretization:**

**Type of Solution:** ILU Preconditioned BiCGSTAB Method

**Type of ILU used:** ILU (0) : Ref. *Iterative Methods for Sparse Linear Systems* by *Y. Saad*

**Boundary Condition:** Dirichlet

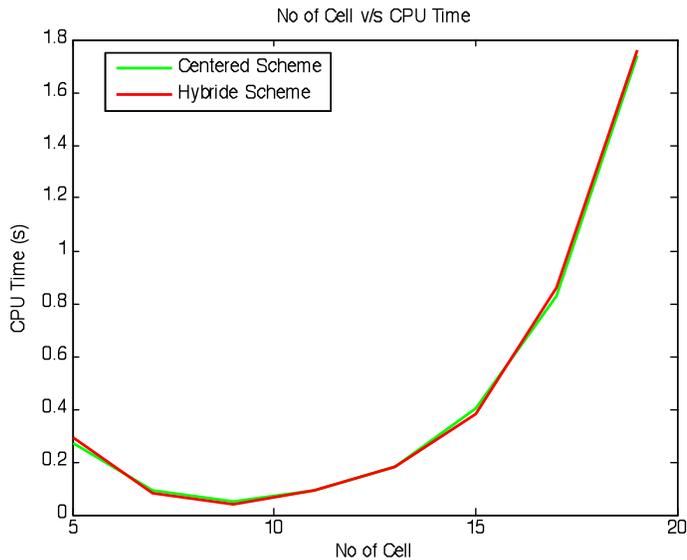

**Comment:**

Time required does not depend on the method of descritization.

**Comparison of Different Method of Discretization:**

Descritization Scheme : Hybrid

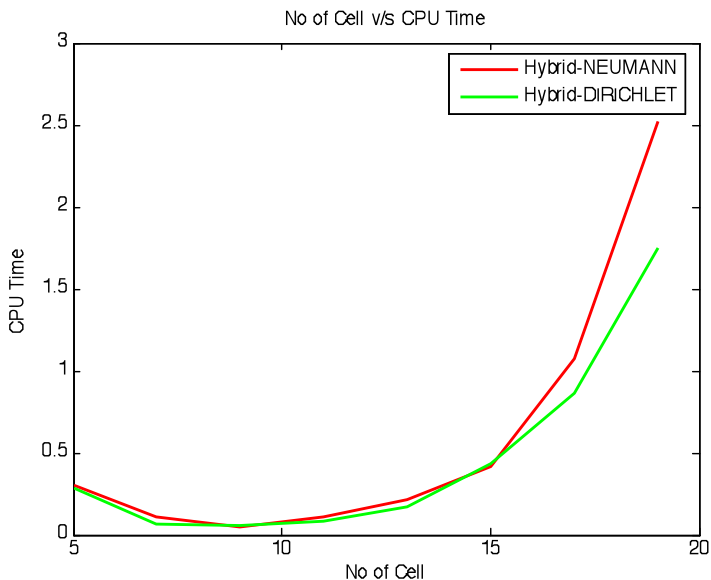

**Comment:**

Time required for the Neumann condition is more as compared to that of its Dirichlet counterpart. According to the authors, the probable reason is that for the case of Neummann condition, thesolution at the nodes where Dirichlet is specified are not known.Hence it takes more time to compute those unknowns.

# Conclusive Discussion:

- Among all the methods discussed, Preconditioned BiCGSTAB method plays well as per as CPU time and convergence and other features are considered.

- Large number of cell gives better result. But it takes more time to converge.

- Type of preconditioned used also plays role in the CPU time consumption and the converges of the solution is considered.

- The value of $\varepsilon$ plays a crucial role as depending on the $\varepsilon$, the contribution of the diffusion term will determined.

# Acknowledgement:

The authors of the report are highly indebted to **Prof.Jean Pequit**– the instructor of the course for his outstanding way of teaching throughout the course. It is because of his awesome way of teachings; the authors were able to understand the subject matter very well and have been able to finish the project.

# Appendix

## The Matlab Code

| Matlab Code Name | Matrix_A_b_U.m |
|---|---|
| Brief Description | The function file takes the following input: eps,X,Y,m,n,Ux0,UxX,Uy0,UyY,f and it does the descritization of the continuous equation using the **hybrid scheme** and formulate the linear system: Ax=b using Dirichlet boundary condition. |

```matlab
function [A,b]=Matrix_A_b_U(eps,X,Y,m,n,Ux0,UxX,Uy0,UyY,f)
%---------------------Help File For Matrix_A_b-----------------------
%     Solution of the elliptic PDE
%     eps(u_xx + u_yy)+a(u_x)+b(u_y) = f(x,y), in a rectangular domain
% [0,X]x[0,Y]
%     with Dirichlet boundary conditions.
%     [A,b]=Matrix_A_b(X,Y,m,n,Ux0,UxX,Uy0,UyY,f)
%     OutPut:
%       A      : Matrix for the system of equation to be solved
%                Ax=b
%       b      : Force Vector
%     Input:
%       X,Y    : Lenghts for the domain definition in X and Y direction
%       m,n    : number of intervals for discretization in x and y
%       Ux0,UxX: Boundary conditions at x=0 and x=X
%                Constant value (scalar) or column vector with the
%                prescribed value at each boundary point
%       Uy0,UyY: Boundary conditions at y=0 and y=Y
%                Constant value (scalar) or row vector with the
%                prescribed value at each boundary point
%       f      : Source term. Scalar (constant source term) or
%                matrix with dimension (n-1)*(m-1)
%                (only interior nodes).
%================================================================

% Checking of de dimensions for the input data
% ----------------------------------------------
[f,Ux0,UxX,Uy0,UyY]=checkInputData(n,m,f,Ux0,UxX,Uy0,UyY);
%================================================================
%           CONSTRUCTION OF THE PENTADIAGONAL MATRIX
%================================================================

% ----------------------------------------------------------------
% Initial Data
% ----------------------------------------------------------------
dime = (m-1)*(n-1);
disp(strcat('Matrix dimension = ',num2str(dime)));
Ax = X/m; Ay = Y/n; %step size for both directions
x=[0:Ax:X]';
y=[0:Ay:Y]';
```

```matlab
%--------------------------------------------------------------------------
% Extracting x,y coordinate and creating the a,b vector:
% --------------------------------------------------------------------------
% Creation of the a,b matrices:
a=zeros((n-1),(m-1));
b=zeros((n-1),(m-1));

for j=1:n-1
    for i=1:m-1
        x_cord=i*Ax;
        y_cord=j*Ay;
        
        a(j,i)=1+(x_cord)^2;
        b(j,i)=X*exp(-y_cord);
        
    end
end

% Creation of the a,b vector:
avec=zeros(dime,1);
bvec=zeros(dime,1);

for j=1:(n-1)
    k=(j-1)*(m-1);
    avec(k+1:j*(m-1))=a(j,:);
    bvec(k+1:j*(m-1))=b(j,:);
end

%--------------------------------------------------------------------------
% Defining the coefficient of U
%--------------------------------------------------------------------------
Diag=ones(dime,1);

%Coefficient of U_i_jm1
%----------------------
    T_i_jm1=eps*Ax*Ax*Diag-bvec*Ax*Ax*Ay;

%Coefficient of U_im1_j
%----------------------
    T_im1_j=eps*Ay*Ay*Diag-avec*Ax*Ay*Ay;

%Coefficient of U_i_j
%----------------------
  T_i_j=-2*eps*Ay*Ay*Diag-2*eps*Diag*Ax*Ax+avec*Ax*Ay*Ay+bvec*Ax*Ax*Ay;

%Coefficient of U_ip1_j
%----------------------
T_ip1_j=eps*Ay*Ay*Diag;

%Coefficient of U_i_jp1
%----------------------
T_i_jp1=eps*Ax*Ax*Diag;
```

```matlab
%--------------------------------------------------------------------
% Creation of the Matrix A:
%--------------------------------------------------------------------
A=zeros(dime,dime);

% Main Diagonal
% -------------
for i=1:dime
    A(i,i)=T_i_j(i);
end

% Up Diagonal
% ----------

for i=1:(dime-1)
    A(i,i+1)=T_ip1_j(i);
end
for i=1:(n-2)
    A(i*(m-1),i*(m-1)+1)=0;
end

% Down Diagonal
% ------------
for i=2:dime
    A(i,i-1)=T_im1_j(i);
end
for i=1:(n-2)
    A(i*(m-1)+1,i*(m-1))=0;
end

% Upper Term
% ----------

for i=1:(n-2)*(m-1)
    A(i,i+(m-1))=T_i_jp1(i);
end

% Lower Term
% ----------
for i=1:(n-2)*(m-1)
    A((m-1)+i,i)=T_i_jm1(i+(m-1));
end

% Plot of the matrix non-zero entries
% -----------------------------------
%   figure
 spy(A)
 title('Plot of the matrix non-zero entries')

%--------------------------------------------------------------------
% Construction of b matrix
```

```matlab
%----------------------------------------------------------------
b = zeros(dime,1);
%Source term
for j = 1:n-1
   k = (j-1)*(m-1);
   b(k+1:k+m-1) = (Ax*Ax)*(Ay*Ay)*f(j,:);
end

%Boundary conditions (prescrived values)
b(1:m-1) = b(1:m-1)-T_i_jm1(1:m-1).*Uy0(2:m);
k = (n-2)*(m-1);
b(k+1:k+m-1) = b(k+1:k+m-1)-T_i_jp1(k+1:k+m-1).*UyY(2:m);
b(1:m-1:dime) = b(1:m-1:dime)-T_im1_j(1:m-1:dime).*Ux0(2:n);
b(m-1:m-1:dime) = b(m-1:m-1:dime)-T_ip1_j(m-1:m-1:dime).*UxX(2:n);

%%%%%%%%%%%%%%%%%%%%%%%%%%%%%%%%%%%%%%%%%%%%%%%%%%%%%%%%%%%%%%%%
%Routine to check the dimensions of the input data
% Checking of de dimensions for the input data
%%%%%%%%%%%%%%%%%%%%%%%%%%%%%%%%%%%%%%%%%%%%%%%%%%%%%%%%%%%%%%%%
function [rf,rUx0,rUxX,rUy0,rUyY] = checkInputData(n,m,f,Ux0,UxX,Uy0,UyY)

[n0 m0] = size(f);
if((n0==1)&(m0==1))   f = f*ones(n-1,m-1); %Constant source term
elseif((n0~=n-1) | (m0~=m-1))
    error ('Error: wrong dimensions for the source term')
end
[n0 m0] = size(Ux0);
if((n0==1)&(m0==1))   Ux0 = Ux0*ones(n+1,1); %Constant BC
elseif((n0~=n+1) | (m0~=1))
    error ('Error: wrong dimensions for the BC at x=0')
end
[n0 m0] = size(UxX);
if((n0==1)&(m0==1))   UxX = UxX*ones(n+1,1);
elseif((n0~=n+1) | (m0~=1))
    error ('Error: wrong dimensions for the BC at x=a')
end
[n0 m0] = size(Uy0);
if((n0==1)&(m0==1))   Uy0 = Uy0*ones(m+1,1);
elseif((n0~=m+1) | (m0~=1))
    error ('Error: wrong dimensions for the BC at y=0')
end
[n0 m0] = size(UyY);
if((n0==1)&(m0==1))   UyY = UyY*ones(m+1,1);
elseif((n0~=m+1) | (m0~=1))
    error ('Error: wrong dimensions for the BC at y=b')
end

%Output assignement
rf = f;
rUx0=Ux0;
rUxX=UxX;
rUy0=Uy0;
```

```
rUyY=UyY;
%%%%%%%%%%%%%%%%%%%%%%%%%%%%%%%%%%%%%%%%%%%%%%%%%%%%%%%%%%%%%%%%%%%%%%%%
```

| **Matlab Code Name** | Matrix_A_b_C.m |
|---|---|
| **Brief Description** | The function file takes the following input: eps,X,Y,m,n,Ux0,UxX,Uy0,UyY,f and it does the descritization of the continuous equation using the **centered scheme** and formulate the linear system: Ax=b Dirichlet boundary condition. |

```
function [A,b]=Matrix_A_b_C(eps,X,Y,m,n,Ux0,UxX,Uy0,UyY,f)
%--------------------Help File For Matrix_A_b---------------------------
-
%     Solution of the elliptic PDE
%     eps(u_xx + u_yy)+a(u_x)+b(u_y) = f(x,y), in a rectangular domain
[0,X]x[0,Y]
%     with Dirichlet boundary conditions.
%     [A,b]=Matrix_A_b(X,Y,m,n,Ux0,UxX,Uy0,UyY,f)
%     OutPut:
%        A       : Matrix for the system of equation to be solved
%                  Ax=b
%        b       : Force Vector
%     Input:
%        X,Y     : Lenghts for the domain definition in X and Y direction
%        m,n     : number of intervals for discretization in x and y
%        Ux0,UxX: Boundary conditions at x=0 and x=X
%                  Constant value (scalar) or column vector with the
%                  prescribed value at each boundary point
%        Uy0,UyY: Boundary conditions at y=0 and y=Y
%                  Constant value (scalar) or row vector with the
%                  prescribed value at each boundary point
%        f       : Source term. Scalar (constant source term) or
%                  matrix with dimension (n-1)*(m-1)
%                  (only interior nodes).
%=======================================================================

% Checking of de dimensions for the input data
% ------------------------------------------------
[f,Ux0,UxX,Uy0,UyY]=checkInputData(n,m,f,Ux0,UxX,Uy0,UyY);
%=======================================================================
%            CONSTRUCTION OF THE PENTADIAGONAL MATRIX
%=======================================================================

% ----------------------------------------------------------------------
% Initial Data
% ----------------------------------------------------------------------
dime = (m-1)*(n-1);
disp(strcat('Matrix dimension = ',num2str(dime)));
Ax = X/m; Ay = Y/n; %step size for both directions
x=[0:Ax:X]';
y=[0:Ay:Y]';
%-----------------------------------------------------------------------
```

```matlab
% Extracting x,y coordinate and creating the a,b vector:
% ----------------------------------------------------------------
% Creation of the a,b matrices:
a=zeros((n-1),(m-1));
b=zeros((n-1),(m-1));

for j=1:n-1
    for i=1:m-1
        x_cord=i*Ax;
        y_cord=j*Ay;

        a(j,i)=1+(x_cord)^2;
        b(j,i)=X*exp(-y_cord);

    end
end

% Creation of the a,b vector:
avec=zeros(dime,1);
bvec=zeros(dime,1);

for j=1:(n-1)
    k=(j-1)*(m-1);
    avec(k+1:j*(m-1))=a(j,:);
    bvec(k+1:j*(m-1))=b(j,:);
end

%----------------------------------------------------------------
% Defining the coefficient of U
%----------------------------------------------------------------
Diag=ones(dime,1);

%Coefficient of U_i_jm1
%-----------------------
   T_i_jm1=eps*Ax*Ax*Diag-bvec*0.50*Ax*Ax*Ay;

%Coefficient of U_im1_j
%-----------------------
   T_im1_j=eps*Ay*Ay*Diag-avec*0.50*Ax*Ay*Ay;

%Coefficient of U_i_j
%-----------------------
  T_i_j=-2*eps*Ay*Ay*Diag-2*eps*Diag*Ax*Ax;

%Coefficient of U_ip1_j
%-----------------------
T_ip1_j=eps*Ay*Ay*Diag+avec*0.50*Ax*Ay*Ay;

%Coefficient of U_i_jp1
%-----------------------
T_i_jp1=eps*Ax*Ax*Diag+bvec*0.50*Ax*Ax*Ay;
```

```matlab
%-----------------------------------------------------------------------
% Creation of the Matrix A:
%-----------------------------------------------------------------------
A=zeros(dime,dime);

% Main Diagonal
% --------------
for i=1:dime
    A(i,i)=T_i_j(i);
end

% Up Diagonal
% ----------

for i=1:(dime-1)
    A(i,i+1)=T_ip1_j(i);
end
for i=1:(n-2)
    A(i*(m-1),i*(m-1)+1)=0;
end

% Down Diagonal
% ------------
for i=2:dime
    A(i,i-1)=T_im1_j(i);
end
for i=1:(n-2)
    A(i*(m-1)+1,i*(m-1))=0;
end

% Upper Term
% ----------

for i=1:(n-2)*(m-1)
    A(i,i+(m-1))=T_i_jp1(i);
end

% Lower Term
% ----------
for i=1:(n-2)*(m-1)
    A((m-1)+i,i)=T_i_jm1(i+(m-1));
end

% Plot of the matrix non-zero entries
% -----------------------------------------
%   figure
 spy(A)
 title('Plot of the matrix non-zero entries')

%-----------------------------------------------------------------------
% Construction of b matrix
%-----------------------------------------------------------------------
```

```matlab
b = zeros(dime,1);
%Source term
for j = 1:n-1
   k = (j-1)*(m-1);
   b(k+1:k+m-1) = (Ax*Ax)*(Ay*Ay)*f(j,:);
end

%Boundary conditions (prescrived values)
b(1:m-1) = b(1:m-1)-T_i_jm1(1:m-1).*Uy0(2:m);
k = (n-2)*(m-1);
b(k+1:k+m-1) = b(k+1:k+m-1)-T_i_jp1(k+1:k+m-1).*UyY(2:m);
b(1:m-1:dime) = b(1:m-1:dime)-T_im1_j(1:m-1:dime).*Ux0(2:n);
b(m-1:m-1:dime) = b(m-1:m-1:dime)-T_ip1_j(m-1:m-1:dime).*UxX(2:n);

%%%%%%%%%%%%%%%%%%%%%%%%%%%%%%%%%%%%%%%%%%%%%%%%%%%%%%%%
%Routine to check the dimensions of the input data
% Checking of de dimensions for the input data
%%%%%%%%%%%%%%%%%%%%%%%%%%%%%%%%%%%%%%%%%%%%%%%%%%%%%%%%
function [rf,rUx0,rUxX,rUy0,rUyY] = checkInputData(n,m,f,Ux0,UxX,Uy0,UyY)

[n0 m0] = size(f);
if((n0==1)&(m0==1))  f = f*ones(n-1,m-1); %Constant source term
elseif((n0~=n-1) | (m0~=m-1))
    error ('Error: wrong dimensions for the source term')
end
[n0 m0] = size(Ux0);
if((n0==1)&(m0==1))  Ux0 = Ux0*ones(n+1,1); %Constant BC
elseif((n0~=n+1) | (m0~=1))
    error ('Error: wrong dimensions for the BC at x=0')
end
[n0 m0] = size(UxX);
if((n0==1)&(m0==1))  UxX = UxX*ones(n+1,1);
elseif((n0~=n+1) | (m0~=1))
    error ('Error: wrong dimensions for the BC at x=a')
end
[n0 m0] = size(Uy0);
if((n0==1)&(m0==1))  Uy0 = Uy0*ones(m+1,1);
elseif((n0~=m+1) | (m0~=1))
    error ('Error: wrong dimensions for the BC at y=0')
end
[n0 m0] = size(UyY);
if((n0==1)&(m0==1))  UyY = UyY*ones(m+1,1);
elseif((n0~=m+1) | (m0~=1))
    error ('Error: wrong dimensions for the BC at y=b')
end

%Output assignement
rf = f;
rUx0=Ux0;
rUxX=UxX;
rUy0=Uy0;
rUyY=UyY;
```

| **Matlab Code Name** | ILUdecomposition2.m |
|---|---|
| **Brief Description** | The function file takes A and b matrix as input and performs the ILU(0) decomposition based on the algorithm described in Y. Saad book. |

```matlab
function [Lo,Up,R] = ILUdecomposition2(A)

% LU decomposition of matrix A; returns A = [L\U].
% disp('Original A:');
% A;
A_original=A;
n = size(A,1);
for k = 1:n-1
for i = k+1:n
if A(i,k) ~= 0.0
lambda = A(i,k)/A(k,k);
A(i,k+1:n) = A(i,k+1:n) - lambda*A(k,k+1:n);
A(i,k) = lambda;
end
end
end

for i=1:n
    for j=1:n
    if A_original(i,j)==0;
    A(i,j)=0;
    end
    end
end

% disp('Modified A=[L\U]:');
% A;

n=length(A);
Lo=zeros(n);
Up=zeros(n);

for i=1:n
    for j=1:i
        Lo(i,j)=A(i,j);
        Lo(i,i)=1;
    end
    for j=i:n
        Up(i,j)=A(i,j);
    end
end

%
% disp('Lo:');
% Lo
```

```matlab
% disp('Up:');
% Up
A_new=Lo*Up;
R=A_new-A;
```

| Matlab Code Name | ILUdecomposition.m |
|---|---|
| Brief Description | The function file takes A and b matrix as input and performs the ILU decomposition based on the algorithm mentioned by the instructor Prof.J.Piquet |

```matlab
function [Lo,Up,R] = ILUdecomposition(A)

n=length(A);
Lo=zeros(n);
Up=zeros(n);

for i=1:n
    for j=1:i
        Lo(i,j)=A(i,j);
        Lo(i,i)=1;
    end
    for j=i:n
        Up(i,j)=A(i,j);
    end
end
Up;
Lo;
A_new=Lo*Up;
R=A_new-A;
```

| Matlab Code Name | BiCGSTAB2.m |
|---|---|
| Brief Description | The function file performs the preconditioned BiCGSTAB algorithm taking A,b,Lower, Upper Triangular and the residual matrix as input. |

```matlab
function [x,residue_mat,iter_mat,amp_mat]=BiCGSTAB2(Lo,Up,R,A,b)

tolerence=1e-5;  % Tolerence Limit
residue=1        % Initial error
iter=0;          % Initial Iteration

residue_mat=[];
iter_mat=[];

% Compute the residueue first

% Initial guess on x:
la=length(A);
```

```matlab
x=ones(la,1)*10;
%residueue:
% b=inv(M)*b;
r=b-A*x;
initial_norm=norm(r)
%R_Tilda:
r_tilda=r;

while residue>tolerence
    iter = iter+1
    rho_1=dot(r_tilda,r);
    if rho_1 ==0
       fprintf('ERROR:\n')
    break;
    end;
    if iter==1
       p=r;
    else
       beta=(rho_1/rho_2)*(alph/omega);
       p = r+beta*(p-omega*v);
    end

    % Solve the system Mp_hat=p, v=Ap_hat:
     M=Lo*Up;
%      M=inv(M)*A;
    p_hat= inv(M)*p;
    v=A*p_hat;

    % Carry on:
    alph=(rho_1/dot(r_tilda,v));
    s=r-alph*v;
%      residue_s=norm(s)/norm(b)
     residue_s=norm(s);
    if residue_s<tolerence
        disp('done')
        x=x+alph*p_hat;
        break
    end

   % Solve Mz=s ; t=Az , z=s_hat
   s_hat=inv(M)*s;
   t=A*s_hat;

   omega=dot(t,s)/dot(t,t);
   x=x+alph*p_hat+omega*s_hat;
   r=s-omega*t;

   rho_2=rho_1;

   % Check for Tolerence
```

```matlab
%     residue=norm(r)/norm(b);
    residue=norm(r)

    residue_mat=[residue_mat residue];
    iter_mat=[iter_mat iter];

    if residue<tolerence
        break
    end

    if omega==0
        break
    end
end
amp_mat=zeros(length(residue_mat)-1,1);
for i=2:length(residue_mat)
    amp_mat(i-1)=residue_mat(i)/residue_mat(i-1);
end
```

%%%%%%%%%%%%%%%%%%%%%%%%%%%%%%%%%%%%%%%%%%%%%%%%%%%

| Matlab Code Name | BiCGSTAB2_nopre.m |
|---|---|
| Brief Description | The function file performs the NO preconditioned BiCGSTAB algorithm taking A,b as input. |

```matlab
function [x,residue_mat,iter_mat,amp_mat]=BiCGSTAB2_nopre(A,b)

tolerence=1e-20;  % Tolerence Limit
residue=1         % Initial error
iter=0;           % Initial Iteration

residue_mat=[];
iter_mat=[];

% Compute the residueue first

% Initial guess on x:
la=length(A);

x=ones(la,1)*100000;
%residueue:
r=b-A*x;
initial_norm=norm(r)
%R_Tilda:
r_tilda=r;

while residue>tolerence
```

```matlab
    iter = iter+1
    rho_1=dot(r_tilda,r);
   if rho_1 ==0
       fprintf('ERROR:\n')
     break;
     end;
     if iter==1
         p=r;
     else
         beta=(rho_1/rho_2)*(alph/omega);
         p = r+beta*(p-omega*v);
     end

    % Solve the system Mp_hat=p, v=Ap_hat:
    M=eye(length(A));
    p_hat= inv(M)*p;
    v=A*p_hat;

    % Carry on:
    alph=(rho_1/dot(r_tilda,v));
    s=r-alph*v;
%      residue_s=norm(s)/norm(b)
     residue_s=norm(s);
    if residue_s<tolerence
        disp('done')
        x=x+alph*p_hat;
        break
    end

   % Solve Mz=s ; t=Az , z=s_hat
   s_hat=inv(M)*s;
   t=A*s_hat;

   omega=dot(t,s)/dot(t,t);
   x=x+alph*p_hat+omega*s_hat;
   r=s-omega*t;

   rho_2=rho_1;

   % Check for Tolerence

%     residue=norm(r)/norm(b);
   residue=norm(r)

   residue_mat=[residue_mat residue];
   iter_mat=[iter_mat iter];

   if residue<tolerence
       break
   end

   if omega==0
```

```matlab
            break
        end
    end
end
amp_mat=zeros(length(residue_mat)-1,1);
for i=2:length(residue_mat)
    amp_mat(i-1)=residue_mat(i)/residue_mat(i-1);
end
```

%%%%%%%%%%%%%%%%%%%%%%%%%%%%%%%%%%%%%%%%%%%%%%%%%%%

| **Matlab Code Name** | No_Acl_Sol.m |
|---|---|
| **Brief Description** | The function file performs the iterative solution on the LU decomposed system. It does not use the acceleration method |

```matlab
function [x_true,resid_mat_noac,iter_mat,amp_mat]=No_Acl_sol(Lo,Up,R,b)

tolerence=1e-5;   % Tolerence Limit
residue=1;        % Initial error
iter=0;           % Initial Iteration

resid_mat_noac=[];
iter_mat=[];

% Compute the residueue first

% Initial guess on x:
la=length(b);

x_guess=rand(la,1);

A=Lo*Up;

while residue>tolerence
    iter = iter+1

%    x_true=(inv(Lo*Up))*R*x_guess+(inv(Lo*Up))*b;

%---- Solving using LU decompositio -------
   b_new=R*x_guess+b;
%   A = LUdec(A);
%   x_true = LUsol(A,b_new);
   x_true = inv(A)*b_new;
   error=(x_guess-x_true);
%-----------------------------------------
   residue=norm(error);
```

```matlab
        resid_mat_noac=[resid_mat_noac residue];
        iter_mat=[iter_mat iter];

        if residue<tolerence
            break
        end

    x_guess=x_true;
end
%%%%%%%%%%%%%%%%%%%%%%%%%%%%%%%%%%%%%%%%%%%%%%%%%%%%
amp_mat=zeros(length(resid_mat_noac)-1,1);
for i=2:length(resid_mat_noac)
    amp_mat(i-1)=resid_mat_noac(i)/resid_mat_noac(i-1);
end
%%%%%%%%%%%%%%%%%%%%%%%%%%%%%%%%%%%%%%%%%%%%%%%%%%%%
```

| **Matlab Code Name** | main_2.m |
|---|---|
| **Brief Description** | The main file performs the solution using the function file Matrix_A_b_U(for the hybrid case), Matrix_A_b_C(for the centerd case),ILUdecomposition2.m (for I LU(0), ), ILUdecomposition.m (for I LU(direct), ),BiCGSTAB.m(for preconditioned acceleration), BiCGSTAB_no_pre.m(for NO preconditioned acceleration),No_Acl_Sol.m (for solution without acceleration) for the dirichlet condition |

```matlab
%%%%%%%%%%%%%%%%%%%%%%%%%%%%%%%%%%%%%%%%%%%%%%%%%%%%%%%%
%                                                      %
%   SOLUTION OF THE ADVECTION-DIFFUSION EQUATION       %
%      eps(U_xx + U_yy)+a*U_x+b*U_y = f(x,y)           %
%      IN A REXTANGULAR DOMAIN (0,X) X (0,Y)           %
%                                                      %
%%%%%%%%%%%%%%%%%%%%%%%%%%%%%%%%%%%%%%%%%%%%%%%%%%%%%%%%
clear all
close all
clc
% DEFINITION OF THE DOMAIN AND THE DISCRETIZACION
% ---------------------------------------------------
eps=4;
X=4; Y=4;
m=input(' Number of intervals for x direction: ');
Ax=X/m; Ay=Ax; n = round(m*Y/X); %Step size for discretization (same in both directions)
x=[0:Ax:X]';
y=[0:Ay:Y]';
%----------------------------------------------------------------------
% Defining the boundary condition:
% ---------------------------------
Ux0 = (1-exp(-y/Y)).*y;
UxX = exp(-1).*(1-exp(-y/Y)).*y;
Uy0 = 0;
```

```matlab
UyY = exp(-x/X).*(1-exp(-1))*Y;
% --------------------------------
% Defining the sorce term and Uanalytical:
% ------------------------
 f=zeros(n-1,m-1);
 Uanalytical=zeros(n-1,m-1);

for j=1:n-1
    for i=1:m-1
        x_cord=i*Ax;
        y_cord=j*Ay;

        T1_eps=eps*((y_cord/X^2).*exp(-x_cord/X)+(2/Y-y_cord/Y^2-y_cord/X^2).*exp(-(x_cord/X+y_cord/Y)));
        T2=exp(-x_cord/X).*(X*exp(-y_cord)-(y_cord.*(1+x_cord.^2)/X));
        T3=exp(-x_cord/X-y_cord/Y).*((y_cord.*(1+x_cord.^2)/X)-X*exp(-y_cord)+(X/Y)*y_cord*exp(-y_cord));

        f(j,i)=T1_eps+T2+T3;
        Uanalytical(j,i) = exp(-x_cord/X).*(1-exp(-y_cord/Y)).*y_cord;
    end
end

%-----------------------------------------------------------------------
%              Computation of the mesh Reynolds number
%-----------------------------------------------------------------------
a_mat=zeros(n+1,m+1);
b_mat=zeros(n+1,m+1);

for j=1:n+1
    for i=1:m+1
        x_cord=(i-1)*Ax;
        y_cord=(j-1)*Ay;

        a_mat(j,i)=(1+(x_cord)^2)*(Ax/eps);
        b_mat(j,i)=(X*exp(-y_cord))*(Ax/eps);
    end
end
%-----------------------------------------------------------------------
% SOLUTION OF THE BOUNDARY PROBLEM
% using the routine Matrix_A_b_2
[A,b]=Matrix_A_b_U(eps,X,Y,m,n,Ux0,UxX,Uy0,UyY,f);

%----------------------------------------------------
% % Solution by direct inverse
% % -------------------------
%   sol=inv(A)*b;

%--------------------------------------------------
% % Solve using ILU and BiCGSTAB
% % -----------------------------------------------
```

```matlab
% A_ori=A;
    [Lo,Up,R] = ILUdecomposition(A);
% A=A_ori;
%    [sol,residue_mat,iter_mat,amp_mat]=BiCGSTAB2(Lo,Up,R,A,b);
% sol;
%   [sol,residue_mat,iter_mat,amp_mat]=BiCGSTAB2_nopre(A,b);
    [sol,resid_mat_noac,iter_mat,amp_mat]=No_Acl_sol(Lo,Up,R,b);
%---------------------------------------------------

%---------------------------
%POSTPROCESS OF THE SOLUTION
%Store the vector solution in a matrix for plot:
%-------------------------------------------------
U = zeros(n+1,m+1);
%Prescribed values at boundary points
U(1,:)   = Uy0'; U(n+1,:) = UyY';
U(:,1)   = Ux0;  U(:,m+1) = UxX;
%Interior points
for j = 1:n-1
    k = (j-1)*(m-1);
     U(j+1,2:m)=sol(k+[1:m-1])';
end

%%%%%%%%%%%%%%%%%%%%%%%%%%%%%%%%%%%%%%%%%%%%%%%%%%%%%
% U_Analytical_Matrix
U_Analytical_Matrix = zeros(n+1,m+1);

%Prescribed values at boundary points
U_Analytical_Matrix(1,:)   = Uy0'; U_Analytical_Matrix(n+1,:) = UyY';
U_Analytical_Matrix(:,1)   = Ux0;  U_Analytical_Matrix(:,m+1) = UxX;

%Interior points
for j = 1:n-1
    k = (j-1)*(m-1);
    U_Analytical_Matrix(j+1,2:m)=Uanalytical(j,:);
end

%%%%%%%%%%%%%%%%%%%%% Error Plot %%%%%%%%%%%%%%%%%%%%%%
Error_mat=abs(U-U_Analytical_Matrix);

surf(x,y,Error_mat);
xlabel('x')
ylabel('y')
zlabel('Error(x,y)')
title('Error Plot for the solution using ILU(0)preconditioned BiCGSTAB')
%%%%%%%%%%%%%%%%%%%%%%%%%%%%%%%%%%%%%%%%%%%%%%%%%%%%%%
%Plot of the solution
figure
clf; %Clear of the Figure
surf(x,y,U)  %Plot of the solution surface
xlabel('x')
ylabel('y')
```

```matlab
zlabel('U(x,y)')
title('Numerical solution using ILU(0)preconditioned BiCGSTAB')
% axis equal

% figure
% surf(x,y,U_Analytical_Matrix) %Plot of the solution surface
% % axis equal
% title('Analytical plot')
% figure
% contour(x,y,U,20) %Contour plot of the solution
% axis equal
% title('Contour plot')
figure
plot(iter_mat,residue_mat/residue_mat(1))
xlabel('No of Iteration')
ylabel('Norm of Residue')
title('Residual Reduction Plot')

%  figure
%  plot(iter_mat,resid_mat_noac/resid_mat_noac(1))
% xlabel('No of Iteration')
% ylabel('Residue')
% title('Residual Reduction Plot')

figure
plot(iter_mat(2:length(iter_mat)),amp_mat)
xlabel('No of Iteration')
ylabel('Ratio of Norm of Residue(Amplification Factor)')
title('Amplification Factor Plot')
% title('Contour plot')
% figure
% plot(iter_mat,resid_mat_noac)

% figure
% surf(x,y,a_mat);
% xlabel('x')
% ylabel('y')
% zlabel('ah/e')
% title('Mesh Reynolds Number Plot')
%
% figure
% surf(x,y,b_mat);
% xlabel('x')
% ylabel('y')
% zlabel('bh/e')
% title('Mesh Reynolds Number Plot')
```

| Matlab Code Name | main_T.m |
|---|---|
| Brief Description | The main file performs the solution for generating the order of convergence plot, and the CPU time vs mesh size plot using the function file Matrix_A_b_U(for the hybrid case), Matrix_A_b_C(for the centerd case),ILUdecomposition2.m (for I LU(0), ), ILUdecomposition.m (for I LU(direct), ),BiCGSTAB.m(for preconditioned acceleration), BiCGSTAB_no_pre.m(for NO preconditioned acceleration),No_Acl_Sol.m (for solution without acceleration) for the dirichlet boundary condition |

```matlab
%%%%%%%%%%%%%%%%%%%%%%%%%%%%%%%%%%%%%%%%%%%%%%%%%%%%%%%%%%%%%
%                                                           %
%   SOLUTION OF THE ADVECTION-DIFFUSION EQUATION            %
%      eps(U_xx + U_yy)+a*U_x+b*U_y = f(x,y)                %
%      IN A REXTANGULAR DOMAIN (0,X) X (0,Y)                %
%                                                           %
%%%%%%%%%%%%%%%%%%%%%%%%%%%%%%%%%%%%%%%%%%%%%%%%%%%%%%%%%%%%%
clear all
close all
clc
% DEFINITION OF THE DOMAIN AND THE DISCRETIZACION
% ------------------------------------------------
eps=10;
X=4; Y=4;
Time_req=[];
no_dis=[];
error_mat=[];

for m=5:2:20;
tic;
Ax=X/m; Ay=Ax; n = round(m*Y/X); %Step size for discretization (same in both directions)
x=[0:Ax:X]';
y=[0:Ay:Y]';
%----------------------------------------------------------------
% Defining the boundary condition:
% ---------------------------------
Ux0 = (1-exp(-y/Y)).*y;
UxX = exp(-1).*(1-exp(-y/Y)).*y;
Uy0 = 0;
UyY = exp(-x/X).*(1-exp(-1))*Y;
% ---------------------------------
% Defining the sorce term and Uanalytical:
% ------------------------
 f=zeros(n-1,m-1);
 Uanalytical=zeros(n-1,m-1);

for j=1:n-1
    for i=1:m-1
        x_cord=i*Ax;
```

```matlab
            y_cord=j*Ay;
            
            T1_eps=eps*((y_cord/X^2).*exp(-x_cord/X)+(2/Y-y_cord/Y^2-y_cord/X^2).*exp(-(x_cord/X+y_cord/Y)));
            T2=exp(-x_cord/X).*(X*exp(-y_cord)-(y_cord.*(1+x_cord.^2)/X));
            T3=exp(-x_cord/X-y_cord/Y).*((y_cord.*(1+x_cord.^2)/X)-X*exp(-y_cord)+(X/Y)*y_cord*exp(-y_cord));
            
            f(j,i)=T1_eps+T2+T3;
            Uanalytical(j,i) = exp(-x_cord/X).*(1-exp(-y_cord/Y)).*y_cord;
    end
end

%-------------------------------------------------------------------------
% SOLUTION OF THE BOUNDARY PROBLEM
% using the routine Matrix_A_b_2
[A,b]=Matrix_A_b_C(eps,X,Y,m,n,Ux0,UxX,Uy0,UyY,f);

%--------------------------------------------------------
% % Solution by direct inverse
% %   --------------------------
%   sol=inv(A)*b;

%---------------------------------------------------------
% % Solve using ILU and BiCGSTAB
% %   ---------------------------------------------------
% A_ori=A;
    [Lo,Up,R] = ILUdecomposition2(A);
% A=A_ori;
    [sol,residue_mat,iter_mat,amp_mat]=BiCGSTAB2(Lo,Up,R,A,b);
% sol;
%    [sol,residue_mat,iter_mat,amp_mat]=BiCGSTAB2_nopre(A,b);
%    [sol,resid_mat_noac,iter_mat,amp_mat]=No_Acl_sol(Lo,Up,R,b);
%-------------------------------------------------

%--------------------------
%POSTPROCESS OF THE SOLUTION
%Store the vector solution in a matrix for plot:
%------------------------------------------------
U = zeros(n+1,m+1);

%Prescribed values at boundary points
U(1,:)   = Uy0'; U(n+1,:) = UyY';
U(:,1)   = Ux0;  U(:,m+1) = UxX;

%Interior points
for j = 1:n-1
    k = (j-1)*(m-1);
     U(j+1,2:m)=sol(k+[1:m-1])';
```

```matlab
    end

    %%%%%%%%%%%%%%%%%%%%%%%%%%%%%%%%%%%%%%%%%%%%%%%%%%%%
    % U_Analytical_Matrix
    U_Analytical_Matrix = zeros(n+1,m+1);

    %Prescribed values at boundary points
    U_Analytical_Matrix(1,:)    = Uy0'; U_Analytical_Matrix(n+1,:) = UyY';
    U_Analytical_Matrix(:,1)    = Ux0;  U_Analytical_Matrix(:,m+1) = UxX;

    %Interior points
    for j = 1:n-1
        k = (j-1)*(m-1);
        U_Analytical_Matrix(j+1,2:m)=Uanalytical(j,:);
    end

    Time_required=toc;
    Error=max(max(abs(U-U_Analytical_Matrix)));

    Time_req=[Time_req Time_required];
    no_dis=[no_dis m];
    error_mat=[error_mat Error];

end

plot(no_dis,Time_req)
xlabel('No of Cell')
ylabel('CPU Time (s)')
title('No of Cell v/s CPU Time')

figure
plot(log(no_dis),log(error_mat))
% loglog(no_dis,error_mat)
  % Least squares approximation
  p=polyfit(log(no_dis),log(error_mat),1);
  x=log(no_dis(3))+0.125;
  y=log(error_mat(3))-0.125;
  text(x,y,num2str(p(1),2))
xlabel('log(No of Cell)')
ylabel('log(Error)')
title('Order of the Scheme')
%------------------------------------------------------------------
% %%%%%%%%%%%%%%%%%% Error Plot %%%%%%%%%%%%%%%%%%%%%%
% Error_mat=abs(U-U_Analytical_Matrix);
%
% surf(x,y,Error_mat);
%
% %%%%%%%%%%%%%%%%%%%%%%%%%%%%%%%%%%%%%%%%%%%%%%%%%%%%%%
% %Plot of the solution
% figure
% clf; %Clear of the Figure
% surf(x,y,U) %Plot of the solution surface
```

```matlab
% % axis equal
% title('Surface plot')
% figure
% surf(x,y,U_Analytical_Matrix) %Plot of the solution surface
% % axis equal
% title('Analytical plot')
% figure
% contour(x,y,U,20) %Contour plot of the solution
% axis equal
% title('Contour plot')
% figure
% plot(iter_mat,residue_mat)
%
```

| **Matlab Code Name** | main_T_Neumann.m |
|---|---|
| **Brief Description** | The main file performs the solution for generating the order of convergence plot, and the CPU time vs mesh size plot using the function file Matrix_A_b_NU(for the hybrid case), Matrix_A_b_NC(for the centerd case),ILUdecomposition2.m (for I LU(0), ), ILUdecomposition.m (for I LU(direct), ),BiCGSTAB.m(for preconditioned acceleration), BiCGSTAB_no_pre.m(for NO preconditioned acceleration),No_Acl_Sol.m (for solution without acceleration) for the Neumann boundary condition |

```matlab
%%%%%%%%%%%%%%%%%%%%%%%%%%%%%%%%%%%%%%%%%%%%%%%%%%%%%%%%%%%
%                                                         %
%   SOLUTION OF THE ADVECTION-DIFFUSION EQUATION          %
%     eps(U_xx + U_yy)+a*U_x+b*U_y = f(x,y)               %
%     IN A REXTANGULAR DOMAIN (0,X) X (0,Y)               %
%                                                         %
%%%%%%%%%%%%%%%%%%%%%%%%%%%%%%%%%%%%%%%%%%%%%%%%%%%%%%%%%%%
clear all
close all
clc
% DEFINITION OF THE DOMAIN AND THE DISCRETIZACION
% --------------------------------------------------
eps=4;
X=4; Y=4;
Time_req=[];
no_dis=[];
error_mat=[];

for m=5:2:20;
tic;

Ax=X/m; Ay=Ax; n = round(m*Y/X); %Step size for discretization (same in both directions)
x=[0:Ax:X]';
y=[0:Ay:Y]';
%-------------------------------------------------------------------
```

```matlab
% Defining the boundary condition:
% ---------------------------------
Ux0 = (1-exp(-y/Y)).*y;
gxX = (-1/X)*exp(-1)*(1-exp(-y/Y)).*y;
Uy0 = 0;
gyY = exp(-x/X).*((Y/Y).*exp(-Y/Y)+(1-exp(-Y/Y)));
% ---------------------------------
% Defining the sorce term and Uanalytical:
% ------------------------
 f=zeros(n-1,m-1);
 Uanalytical=zeros(n,m);

for j=1:n-1
    for i=1:m-1
        x_cord=i*Ax;
        y_cord=j*Ay;

        T1_eps=eps*((y_cord/X^2).*exp(-x_cord/X)+(2/Y-y_cord/Y^2-y_cord/X^2).*exp(-(x_cord/X+y_cord/Y)));
        T2=exp(-x_cord/X).*(X*exp(-y_cord)-(y_cord.*(1+x_cord.^2)/X));
        T3=exp(-x_cord/X-y_cord/Y).*((y_cord.*(1+x_cord.^2)/X)-X*exp(-y_cord)+(X/Y)*y_cord*exp(-y_cord));

        f(j,i)=T1_eps+T2+T3;

    end
end

for j=1:n
    for i=1:m
        x_cord=i*Ax;
        y_cord=j*Ay;
Uanalytical(j,i) = exp(-x_cord/X).*(1-exp(-y_cord/Y)).*y_cord;
    end
end
%-------------------------------------------------------------------
% SOLUTION OF THE BOUNDARY PROBLEM
% using the routine Matrix_A_b_2
[A,b]=Matrix_A_b_NU(eps,X,Y,m,n,Ux0,gxX,Uy0,gyY,f);

%-----------------------------------------------------
% % Solution by direct inverse
% % -------------------------
%   sol=inv(A)*b;

%-----------------------------------------------------
% % Solve using ILU and BiCGSTAB
% % -------------------------------------------------
 [Lo,Up,R] = ILUdecomposition2(A);

 [sol,residue_mat,iter_mat,amp_mat]=BiCGSTAB2(Lo,Up,R,A,b);
```

```matlab
    %[sol,residue_mat,iter_mat,amp_mat]=BiCGSTAB2_nopre(A,b);
    %  [sol,resid_mat_noac,iter_mat,amp_mat]=No_Acl_sol(Lo,Up,R,b);
    %--------------------------------------------------

    %----------------------------
    %POSTPROCESS OF THE SOLUTION
    %Store the vector solution in a matrix for plot:
    %------------------------------------------------
    U = zeros(n+1,m+1);
    %Prescribed values at boundary points
    U(1,:)   = Uy0';
    U(:,1)   = Ux0;
    %Interior points
    for j = 1:n-1
        k = (j-1)*(m-1);
         U(j+1,2:m)=sol(k+[1:m-1])';
    end

    U(2:n,m+1)=U(2:n,m)+Ax*gxX(2:n);
    U(n+1,2:m)=U(n,2:m)+Ay.*gyY(2:m)';

    % U(1:n-1,m)=U(1:n-1,m-1);
    % U(n,1:m-1)=U(n-1,1:m-1);

    %  U(n+1,m+1)=0.50*((U(n+1,m)+Ax*gxX(n+1))+(U(n,m+1)+Ay*gyY(m+1)));

     U(n+1,m+1)=U(n+1,m)+Ax*gxX(n+1);
    %%%%%%%%%%%%%%%%%%%%%%%%%%%%%%%%%%%%%%%%%%%%%%%%%%%
    % U_Analytical_Matrix
    U_Analytical_Matrix = zeros(n+1,m+1);

    %Prescribed values at boundary points
    U_Analytical_Matrix(1,:)   = Uy0';
    U_Analytical_Matrix(:,1)   = Ux0;

    %Interior points
    for j = 1:n
        k = (j-1)*m;
        U_Analytical_Matrix(j+1,2:m+1)=Uanalytical(j,:);
    end

    Time_required=toc;
    Error=max(max(abs(U-U_Analytical_Matrix)));

    Time_req=[Time_req Time_required];
    no_dis=[no_dis m];
    error_mat=[error_mat Error];

end

plot(no_dis,Time_req)
```

```matlab
xlabel('No of Cell')
ylabel('CPU Time')
title('No of Cell v/s CPU Time')

figure
plot(log(no_dis),log(error_mat))
% loglog(no_dis,error_mat)
  % Least squares approximation
  p=polyfit(log(no_dis),log(error_mat),1);
  x=log(no_dis(3))+0.125;
  y=log(error_mat(3))-0.125;
  text(x,y,num2str(p(1),2))
xlabel('log(No of Cell)')
ylabel('log(Error)')
title('Order of the Scheme')
```

| Matlab Code Name | Main_Neumann.m |
|---|---|
| Brief Description | The main file performs the solution using the function file Matrix_A_b_NU(for the hybrid case), Matrix_A_b_NC(for the centerd case),ILUdecomposition2.m (for I LU(0), ), ILUdecomposition.m (for I LU(direct), ),BiCGSTAB.m(for preconditioned acceleration), BiCGSTAB_no_pre.m(for NO preconditioned acceleration),No_Acl_Sol.m (for solution without acceleration) for the Neumann boundary condition |

```matlab
%%%%%%%%%%%%%%%%%%%%%%%%%%%%%%%%%%%%%%%%%%%%%%%%%%%%%%%%%%%%%
%                                                           %
%  SOLUTION OF THE ADVECTION-DIFFUSION EQUATION             %
%     eps(U_xx + U_yy)+a*U_x+b*U_y = f(x,y)                 %
%     IN A REXTANGULAR DOMAIN (0,X) X (0,Y)                 %
%                                                           %
%%%%%%%%%%%%%%%%%%%%%%%%%%%%%%%%%%%%%%%%%%%%%%%%%%%%%%%%%%%%%
clear all
close all
clc
% DEFINITION OF THE DOMAIN AND THE DISCRETIZACION
% --------------------------------------------------
eps=4;
X=4; Y=4;
m=input(' Number of intervals for x direction: ');
Ax=X/m; Ay=Ax; n = round(m*Y/X); %Step size for discretization (same in
both directions)
x=[0:Ax:X]';
y=[0:Ay:Y]';
%-----------------------------------------------------------------------
% Defining the boundary condition:
% -------------------------------
Ux0 = (1-exp(-y/Y)).*y;
gxX = (-1/X)*exp(-1)*(1-exp(-y/Y)).*y;
Uy0 = 0;
gyY = exp(-x/X).*((Y/Y).*exp(-Y/Y)+(1-exp(-Y/Y)));
% -------------------------------
```

```matlab
% Defining the sorce term and Uanalytical:
% ------------------------
 f=zeros(n-1,m-1);
 Uanalytical=zeros(n,m);

for j=1:n-1
    for i=1:m-1
        x_cord=i*Ax;
        y_cord=j*Ay;

        T1_eps=eps*((y_cord/X^2).*exp(-x_cord/X)+(2/Y-y_cord/Y^2-y_cord/X^2).*exp(-(x_cord/X+y_cord/Y)));
        T2=exp(-x_cord/X).*(X*exp(-y_cord)-(y_cord.*(1+x_cord.^2)/X));
        T3=exp(-x_cord/X-y_cord/Y).*((y_cord.*(1+x_cord.^2)/X)-X*exp(-y_cord)+(X/Y)*y_cord*exp(-y_cord));

        f(j,i)=T1_eps+T2+T3;

    end
end

for j=1:n
    for i=1:m
        x_cord=i*Ax;
        y_cord=j*Ay;
Uanalytical(j,i) = exp(-x_cord/X).*(1-exp(-y_cord/Y)).*y_cord;
    end
end
%---------------------------------------------------------------------
% SOLUTION OF THE BOUNDARY PROBLEM
% using the routine Matrix_A_b_2
[A,b]=Matrix_A_b_NU(eps,X,Y,m,n,Ux0,gxX,Uy0,gyY,f);

%--------------------------------------------------------
% % Solution by direct inverse
% % ------------------------
%   sol=inv(A)*b;

%---------------------------------------------------------
% % Solve using ILU and BiCGSTAB
% % ---------------------------------------------------------
   [Lo,Up,R] = ILUdecomposition2(A);

   [sol,residue_mat,iter_mat,amp_mat]=BiCGSTAB2(Lo,Up,R,A,b);

% [sol,residue_mat,iter_mat,amp_mat]=BiCGSTAB2_nopre(A,b);
%    [sol,resid_mat_noac,iter_mat,amp_mat]=No_Acl_sol(Lo,Up,R,b);
%---------------------------------------------------

%---------------------------
%POSTPROCESS OF THE SOLUTION
%Store the vector solution in a matrix for plot:
```

```matlab
%--------------------------------------------------
U = zeros(n+1,m+1);
%Prescribed values at boundary points
U(1,:)    = Uy0';
U(:,1)    = Ux0;
%Interior points
for j = 1:n-1
    k = (j-1)*(m-1);
      U(j+1,2:m)=sol(k+[1:m-1])';
end

U(2:n,m+1)=U(2:n,m)+Ax*gxX(2:n);
U(n+1,2:m)=U(n,2:m)+Ay.*gyY(2:m)';

% U(1:n-1,m)=U(1:n-1,m-1);
% U(n,1:m-1)=U(n-1,1:m-1);

%   U(n+1,m+1)=0.50*((U(n+1,m)+Ax*gxX(n+1))+(U(n,m+1)+Ay*gyY(m+1)));

 U(n+1,m+1)=U(n+1,m)+Ax*gxX(n+1);
%%%%%%%%%%%%%%%%%%%%%%%%%%%%%%%%%%%%%%%%%%%%%%%%%%%
% U_Analytical_Matrix
U_Analytical_Matrix = zeros(n+1,m+1);

%Prescribed values at boundary points
U_Analytical_Matrix(1,:)    = Uy0';
U_Analytical_Matrix(:,1)    = Ux0;

%Interior points
for j = 1:n
    k = (j-1)*m;
    U_Analytical_Matrix(j+1,2:m+1)=Uanalytical(j,:);
end

%%%%%%%%%%%%%%%%%%%% Error Plot %%%%%%%%%%%%%%%%%%%%%
Error_mat=abs(U-U_Analytical_Matrix);

surf(x,y,Error_mat);
xlabel('x')
ylabel('y')
title('Error Plot for the solution using ILU(0)preconditioned BiCGSTAB')
%%%%%%%%%%%%%%%%%%%%%%%%%%%%%%%%%%%%%%%%%%%%%%%%%%%%
%Plot of the solution
figure
clf; %Clear of the Figure
surf(x,y,U)  %Plot of the solution surface
% axis equal
xlabel('x')
ylabel('y')
title('Numerical solution using ILU(0)preconditioned BiCGSTAB')
```

```matlab
% figure
% surf(x,y,U_Analytical_Matrix) %Plot of the solution surface
% % axis equal
% title('Analytical plot')
% figure
% contour(x,y,U,20) %Contour plot of the solution
% axis equal
% title('Contour plot')
figure
plot(iter_mat,residue_mat/residue_mat(1))
xlabel('No of Iteration')
ylabel('Norm of Residue')
title('Residual Reduction Plot')

%  figure
%  plot(iter_mat,resid_mat_noac/resid_mat_noac(1))
% xlabel('No of Iteration')
% ylabel('Residue')
% title('Residual Reduction Plot')

figure
plot(iter_mat(2:length(iter_mat)),amp_mat)
xlabel('No of Iteration')
ylabel('Ratio of Norm of Residue(Amplification Factor)')
title('Amplification Factor Plot')
```

| **Matlab Code Name** | Matrix_A_b_NC.m |
|---|---|
| **Brief Description** | The function file takes the following input: eps,X,Y,m,n,Ux0,gxX,Uy0,gyY,f and it does the descritization of the continuous equation using the **centered scheme** and formulate the linear system: Ax=b  Neumann  boundary condition. |

```matlab
function [A,b]=Matrix_A_b_NC(eps,X,Y,m,n,Ux0,gxX,Uy0,gyY,f)
%--------------------Help File For Matrix_A_b---------------------------
%     Solution of the elliptic PDE
%     eps(u_xx + u_yy)+a(u_x)+b(u_y) = f(x,y), in a rectangular domain [0,X]x[0,Y]
%     with Dirichlet boundary conditions.
%     [A,b]=Matrix_A_b(X,Y,m,n,Ux0,UxX,Uy0,UyY,f)
%     OutPut:
%       A      : Matrix for the system of equation to be solved
%                Ax=b
%       b      : Force Vector
%     Input:
%       X,Y    : Lenghts for the domain definition in X and Y direction
%       m,n    : number of intervals for discretization in x and y
%       Ux0,UxX: Boundary conditions at x=0 and x=X
%                Constant value (scalar) or column vector with the
%                prescribed value at each boundary point
%       Uy0,UyY: Boundary conditions at y=0 and y=Y
```

```matlab
%                      Constant value (scalar) or row vector with the
%                      prescribed value at each boundary point
%         f          : Source term. Scalar (constant source term) or
%                      matrix with dimension (n-1)*(m-1)
%                      (only interior nodes).
%=================================================================

% Checking of de dimensions for the input data
% --------------------------------------------
[f,Ux0,UxX,Uy0,UyY]=checkInputData(n,m,f,Ux0,gxX,Uy0,gyY);
%=================================================================
%              CONSTRUCTION OF THE PENTADIAGONAL MATRIX
%=================================================================

% ----------------------------------------------------------------
% Initial Data
% ----------------------------------------------------------------
dime = (m-1)*(n-1);
disp(strcat('Matrix dimension = ',num2str(dime)));
Ax = X/m; Ay = Y/n; %step size for both directions
x=[0:Ax:X]';
y=[0:Ay:Y]';
%-----------------------------------------------------------------
% Extracting x,y coordinate and creating the a,b vector:
% ----------------------------------------------------------------
% Creation of the a,b matrices:
a=zeros((n-1),(m-1));
b=zeros((n-1),(m-1));

for j=1:n-1
    for i=1:m-1
        x_cord=i*Ax;
        y_cord=j*Ay;

        a(j,i)=1+(x_cord)^2;
        b(j,i)=X*exp(-y_cord);

    end
end

% Creation of the a,b vector:
avec=zeros(dime,1);
bvec=zeros(dime,1);

for j=1:(n-1)
    k=(j-1)*(m-1);
    avec(k+1:j*(m-1))=a(j,:);
    bvec(k+1:j*(m-1))=b(j,:);
end

%-----------------------------------------------------------------
% Defining the coefficient of U
```

```matlab
%--------------------------------------------------------------------
Diag=ones(dime,1);

%Coefficient of U_i_jm1
%-----------------------
    T_i_jm1=eps*Ax*Ax*Diag-bvec*0.50*Ax*Ax*Ay;

%Coefficient of U_im1_j
%-----------------------
    T_im1_j=eps*Ay*Ay*Diag-avec*0.50*Ax*Ay*Ay;

%Coefficient of U_i_j
%-----------------------
   T_i_j=-2*eps*Ay*Ay*Diag-2*eps*Diag*Ax*Ax;

%Coefficient of U_ip1_j
%-----------------------
T_ip1_j=eps*Ay*Ay*Diag+avec*0.50*Ax*Ay*Ay;

%Coefficient of U_i_jp1
%-----------------------
T_i_jp1=eps*Ax*Ax*Diag+bvec*0.50*Ax*Ax*Ay;

%------------------------------------------------------------------------
% Creation of the Matrix A:
%------------------------------------------------------------------------
A=zeros(dime,dime);

% Main Diagonal
% -------------
for i=1:dime
    A(i,i)=T_i_j(i);
end

% Up Diagonal
% ----------

for i=1:(dime-1)
    A(i,i+1)=T_ip1_j(i);
end
for i=1:(n-2)
    A(i*(m-1),i*(m-1)+1)=0;
end

% Down Diagonal
% ------------
for i=2:dime
    A(i,i-1)=T_im1_j(i);
end
for i=1:(n-2)
    A(i*(m-1)+1,i*(m-1))=0;
```

```matlab
end

% Upper Term
% ----------

for i=1:(n-2)*(m-1)
    A(i,i+(m-1))=T_i_jp1(i);
end

% Lower Term
% ----------
for i=1:(n-2)*(m-1)
    A((m-1)+i,i)=T_i_jm1(i+(m-1));
end

%-----------------------------------------------------------------
%  Additional Term for the Neumann condition:

for i=1:(n-1)
    A(i*(m-1),i*(m-1))=A(i*(m-1),i*(m-1))+T_ip1_j(i*(m-1));
end

k=(n-2)*(m-1);
for i=1:(m-1)
    A(k+i,k+i)=A(k+i,k+i)+T_i_jp1(k+i);
end

% Plot of the matrix non-zero entries
% -------------------------------------------
%   figure
  spy(A)
  title('Plot of the matrix non-zero entries')

%-----------------------------------------------------------------
% Construction of b matrix
%-----------------------------------------------------------------
b = zeros(dime,1);
%Source term
for j = 1:n-1
   k = (j-1)*(m-1);
   b(k+1:k+m-1) =  (Ax*Ax)*(Ay*Ay)*f(j,:);
end

%Boundary conditions (prescrived values)
b(1:m-1)  = b(1:m-1)-T_i_jm1(1:m-1).*Uy0(2:m);
k = (n-2)*(m-1);
b(k+1:k+m-1) = b(k+1:k+m-1)-T_i_jp1(k+1:k+m-1).*Ay.*gyY(2:m);
b(1:m-1:dime) = b(1:m-1:dime)-T_im1_j(1:m-1:dime).*Ux0(2:n);
b(m-1:m-1:dime) = b(m-1:m-1:dime)-T_ip1_j(m-1:m-1:dime).*Ax.*gxX(2:n);

%%%%%%%%%%%%%%%%%%%%%%%%%%%%%%%%%%%%%%%%%%%%%%%%%%%%%%%%%%%%
%Routine to check the dimensions of the input data
```

```matlab
% Checking of de dimensions for the input data
%%%%%%%%%%%%%%%%%%%%%%%%%%%%%%%%%%%%%%%%%%%%%%%%%%%%%%
function [rf,rUx0,rgxX,rUy0,rgyY] = checkInputData(n,m,f,Ux0,gxX,Uy0,gyY)

[n0 m0] = size(f);
if((n0==1)&(m0==1))  f = f*ones(n-1,m-1); %Constant source term
elseif((n0~=n-1) | (m0~=m-1))
    error ('Error: wrong dimensions for the source term')
end
[n0 m0] = size(Ux0);
if((n0==1)&(m0==1))  Ux0 = Ux0*ones(n+1,1); %Constant BC
elseif((n0~=n+1) | (m0~=1))
    error ('Error: wrong dimensions for the BC at x=0')
end
[n0 m0] = size(gxX);
if((n0==1)&(m0==1))  gxX = gxX*ones(n+1,1);
elseif((n0~=n+1) | (m0~=1))
    error ('Error: wrong dimensions for the BC at x=a')
end
[n0 m0] = size(Uy0);
if((n0==1)&(m0==1))  Uy0 = Uy0*ones(m+1,1);
elseif((n0~=m+1) | (m0~=1))
    error ('Error: wrong dimensions for the BC at y=0')
end
[n0 m0] = size(gyY);
if((n0==1)&(m0==1))  gyY = gyY*ones(m+1,1);
elseif((n0~=m+1) | (m0~=1))
    error ('Error: wrong dimensions for the BC at y=b')
end

%Output assignement
rf = f;
rUx0=Ux0;
rgxX=gxX;
rUy0=Uy0;
rgyY=gyY;
%%%%%%%%%%%%%%%%%%%%%%%%%%%%%%%%%%%%%%%%%%%%%%%%%%%%%%%%%%%%%%%%%%%%%%%
```

| Matlab Code Name | Matrix_A_b_NU.m |
|---|---|
| Brief Description | The function file takes the following input: eps,X,Y,m,n,Ux0,gxX,Uy0,gyY,f and it does the descritization of the continuous equation using the **upwind scheme** and formulate the linear system: Ax=b  Neumann  boundary condition. |

```matlab
function [A,b]=Matrix_A_b_NU(eps,X,Y,m,n,Ux0,gxX,Uy0,gyY,f)
%--------------------Help File For Matrix_A_b----------------------------
%    Solution of the elliptic PDE
%    eps(u_xx + u_yy)+a(u_x)+b(u_y) = f(x,y), in a rectangular domain [0,X]x[0,Y]
%    with Dirichlet boundary conditions.
%    [A,b]=Matrix_A_b(X,Y,m,n,Ux0,UxX,Uy0,UyY,f)
%    OutPut:
```

```matlab
%          A       : Matrix for the system of equation to be solved
%                   Ax=b
%          b       : Force Vector
%       Input:
%          X,Y     : Lenghts for the domain definition in X and Y direction
%          m,n     : number of intervals for discretization in x and y
%          Ux0,UxX: Boundary conditions at x=0 and x=X
%                   Constant value (scalar) or column vector with the
%                   prescribed value at each boundary point
%          Uy0,UyY: Boundary conditions at y=0 and y=Y
%                   Constant value (scalar) or row vector with the
%                   prescribed value at each boundary point
%          f       : Source term. Scalar (constant source term) or
%                   matrix with dimension (n-1)*(m-1)
%                   (only interior nodes).
%===============================================================

% Checking of de dimensions for the input data
% ---------------------------------------------
[f,Ux0,UxX,Uy0,UyY]=checkInputData(n,m,f,Ux0,gxX,Uy0,gyY);
%===============================================================
%           CONSTRUCTION OF THE PENTADIAGONAL MATRIX
%===============================================================

% --------------------------------------------------------------
% Initial Data
% --------------------------------------------------------------
dime = (m-1)*(n-1);
disp(strcat('Matrix dimension = ',num2str(dime)));
Ax = X/m; Ay = Y/n; %step size for both directions
x=[0:Ax:X]';
y=[0:Ay:Y]';
%--------------------------------------------------------------
% Extracting x,y coordinate and creating the a,b vector:
% --------------------------------------------------------------
% Creation of the a,b matrices:
a=zeros((n-1),(m-1));
b=zeros((n-1),(m-1));

for j=1:n-1
    for i=1:m-1
        x_cord=i*Ax;
        y_cord=j*Ay;
        
        a(j,i)=1+(x_cord)^2;
        b(j,i)=X*exp(-y_cord);
        
    end
end

% Creation of the a,b vector:
avec=zeros(dime,1);
```

```matlab
bvec=zeros(dime,1);

for j=1:(n-1)
    k=(j-1)*(m-1);
    avec(k+1:j*(m-1))=a(j,:);
    bvec(k+1:j*(m-1))=b(j,:);
end

%-----------------------------------------------------------------------
% Defining the coefficient of U
%-----------------------------------------------------------------------
Diag=ones(dime,1);

%Coefficient of U_i_jm1
%-----------------------
    T_i_jm1=eps*Ax*Ax*Diag-bvec*Ax*Ax*Ay;

%Coefficient of U_im1_j
%-----------------------
    T_im1_j=eps*Ay*Ay*Diag-avec*Ax*Ay*Ay;

%Coefficient of U_i_j
%-----------------------
   T_i_j=-2*eps*Ay*Ay*Diag-2*eps*Diag*Ax*Ax+avec*Ax*Ay*Ay+bvec*Ax*Ax*Ay;

%Coefficient of U_ip1_j
%-----------------------
T_ip1_j=eps*Ay*Ay*Diag;

%Coefficient of U_i_jp1
%-----------------------
T_i_jp1=eps*Ax*Ax*Diag;

%-----------------------------------------------------------------------
% Creation of the Matrix A:
%-----------------------------------------------------------------------
A=zeros(dime,dime);

% Main Diagonal
% -------------
for i=1:dime
    A(i,i)=T_i_j(i);
end

% Up Diagonal
% ----------

for i=1:(dime-1)
    A(i,i+1)=T_ip1_j(i);
end
for i=1:(n-2)
    A(i*(m-1),i*(m-1)+1)=0;
```

```matlab
end

% Down Diagonal
% ------------
for i=2:dime
    A(i,i-1)=T_im1_j(i);
end
for i=1:(n-2)
    A(i*(m-1)+1,i*(m-1))=0;
end

% Upper Term
% ----------

for i=1:(n-2)*(m-1)
    A(i,i+(m-1))=T_i_jp1(i);
end

% Lower Term
% ----------
for i=1:(n-2)*(m-1)
    A((m-1)+i,i)=T_i_jm1(i+(m-1));
end

%----------------------------------------------------------------------
%  Additional Term for the Neumann condition:

for i=1:(n-1)
    A(i*(m-1),i*(m-1))=A(i*(m-1),i*(m-1))+T_ip1_j(i*(m-1));
end

k=(n-2)*(m-1);
for i=1:(m-1)
    A(k+i,k+i)=A(k+i,k+i)+T_i_jp1(k+i);
end

% Plot of the matrix non-zero entries
% -----------------------------------
%  figure
 spy(A)
 title('Plot of the matrix non-zero entries')

%----------------------------------------------------------------------
% Construction of b matrix
%----------------------------------------------------------------------
b = zeros(dime,1);
%Source term
for j = 1:n-1
   k = (j-1)*(m-1);
   b(k+1:k+m-1)  =  (Ax*Ax)*(Ay*Ay)*f(j,:);
end
```

```matlab
%Boundary conditions (prescrived values)
b(1:m-1) = b(1:m-1)-T_i_jm1(1:m-1).*Uy0(2:m);
k = (n-2)*(m-1);
b(k+1:k+m-1) = b(k+1:k+m-1)-T_i_jp1(k+1:k+m-1).*Ay.*gyY(2:m);
b(1:m-1:dime) = b(1:m-1:dime)-T_im1_j(1:m-1:dime).*Ux0(2:n);
b(m-1:m-1:dime) = b(m-1:m-1:dime)-T_ip1_j(m-1:m-1:dime).*Ax.*gxX(2:n);

%%%%%%%%%%%%%%%%%%%%%%%%%%%%%%%%%%%%%%%%%%%%%%%%%%%%%%%%
%Routine to check the dimensions of the input data
% Checking of de dimensions for the input data
%%%%%%%%%%%%%%%%%%%%%%%%%%%%%%%%%%%%%%%%%%%%%%%%%%%%%%%%
function [rf,rUx0,rgxX,rUy0,rgyY] = checkInputData(n,m,f,Ux0,gxX,Uy0,gyY)

[n0 m0] = size(f);
if((n0==1)&(m0==1))  f = f*ones(n-1,m-1); %Constant source term
elseif((n0~=n-1) | (m0~=m-1))
    error ('Error: wrong dimensions for the source term')
end
[n0 m0] = size(Ux0);
if((n0==1)&(m0==1))  Ux0 = Ux0*ones(n+1,1); %Constant BC
elseif((n0~=n+1) | (m0~=1))
    error ('Error: wrong dimensions for the BC at x=0')
end
[n0 m0] = size(gxX);
if((n0==1)&(m0==1))  gxX = gxX*ones(n+1,1);
elseif((n0~=n+1) | (m0~=1))
    error ('Error: wrong dimensions for the BC at x=a')
end
[n0 m0] = size(Uy0);
if((n0==1)&(m0==1))  Uy0 = Uy0*ones(m+1,1);
elseif((n0~=m+1) | (m0~=1))
    error ('Error: wrong dimensions for the BC at y=0')
end
[n0 m0] = size(gyY);
if((n0==1)&(m0==1))  gyY = gyY*ones(m+1,1);
elseif((n0~=m+1) | (m0~=1))
    error ('Error: wrong dimensions for the BC at y=b')
end

%Output assignement
rf = f;
rUx0=Ux0;
rgxX=gxX;
rUy0=Uy0;
rgyY=gyY;
%%%%%%%%%%%%%%%%%%%%%%%%%%%%%%%%%%%%%%%%%%%%%%%%%%%%%%%%%%%%%%%%%
```